\documentclass[twocolumn]{aastex631}

\usepackage[utf8]{inputenc}
\usepackage{CJK}
\usepackage[flushleft]{threeparttable}
\usepackage{longtable}
\usepackage{gensymb}
\usepackage{multirow}
\usepackage{array}
\usepackage{amsmath}
\usepackage{mathtools}

\shorttitle{Line velocity of CCSN cores}
\shortauthors{Fang \& Maeda}

\begin{document}
\begin{CJK*}{UTF8}{gbsn}

\title{Inferring the progenitor mass-kinetic energy relation of stripped-envelope core-collapse supernovae from nebular spectroscopy}
\author[0000-0002-1161-9592]{Qiliang Fang (方其亮)}\affiliation{Department of Astronomy, Kyoto University, Kitashirakawa-Oiwake-cho, Sakyo-ku, Kyoto 606-8502, Japan}
\author[0000-0003-2611-7269]{Keiichi Maeda}\affiliation{Department of Astronomy, Kyoto University, Kitashirakawa-Oiwake-cho, Sakyo-ku, Kyoto 606-8502, Japan}

\begin{abstract}
The relation between the progenitor mass and the kinetic energy of the explosion is a key toward revealing the explosion mechanism of stripped-envelope (SE) core-collapse (CC) supernovae (SNe). Here, we present a method to derive this relation using the nebular spectra of SESNe, based on the correlation between the [O~I]/[Ca~II], which is an indicator of the progenitor mass, and the width of [O~I], which measures the expansion velocity of the oxygen-rich material. To explain the correlation, the kinetic energy ($E_{\rm K}$) is required to be positively correlated with the progenitor mass as represented by the CO core mass ($M_{\rm CO}$). We demonstrate that SNe IIb/Ib and SNe Ic/Ic-BL follow the same $M_{\rm CO}$-$E_{\rm K}$ scaling relation, which suggests the helium-rich and helium-deficient SNe share the same explosion mechanism. The $M_{\rm CO}$-$E_{\rm K}$ relation derived in this work is compared with the ones from early phase observations. The results are largely in good agreement. Combined with early phase observation, the method presented in this work provides a chance to scan through the ejecta from the outermost region to the dense inner core, which is important to reveal the global properties of the ejecta and constrain the explosion mechanism of core-collapse supernovae.  
\end{abstract}
 


\section{INTRODUCTION}
Core-collapse supernovae (CCSNe) mark the final stage of the evolution of a massive star (zero-age main-sequence mass larger than 8$M_{\odot}$). The explosion mechanism of this catastrophic event is yet to be clarified. How the properties of the explosion process depends on those of the progenitor is an important open problem in astronomy.

CCSNe are diverse in observable signatures, leading to classification into different subtypes. Type II supernovae (SNe II) show strong hydrogen features in their optical spectra. CCSNe lacking permanent hydrogen signatures are classified as stripped-envelope supernovae (SESNe). Among them, type Ib SNe (SNe Ib) do not exhibit hydrogen features, but their spectra are dominated by helium features. The spectra of type Ic SNe (SNe Ic) lack both hydrogen and helium features. Type IIb SNe (SNe IIb) are transitional events between SNe II and Ib; SNe IIb initially show strong hydrogen signatures, but their spectra eventually resemble to those of SNe Ib as the ejecta continue to evolve. SNe Ic can be further classified as normal SNe Ic and broad line SNe Ic (SNe Ic-BL). The latter type is characterized by the broad absorption features and its (occasional) association with a gamma-ray burst (\citealt{galama98,hjorth03,woosley06}). The readers are refered to \citet{filippenko97}, \citet{galyam17} and \citet{modjaz19} for the classification of SNe. The lack of hydrogen (or helium) in the spectra of SESNe indicates that the hydrogen-rich envelope (or the helium-rich layer) has been stripped away before the explosion. Several channels may be responsible for the pre-SN mass loss, including binary interaction, stellar wind, or a combination of both (\citealt{heger03, sana12, groh13, smith14, yoon15, fang19}).

Understanding how the explosion process is dependent on the state of the progenitor is a key toward revealing the explosion mechanism of CCSNe. We thus need independent measurements of the progenitor properties and the explosion parameters. The relation between the two basic parameters, i.e., the progenitor mass and the amount of the kinetic energy, is particularly important but not established. The main difficulty comes from mapping the observables to these two quantities. The pre-explosion image, which allows one to directly identify the progenitor (therefore provides a robust measurement of the mass), is only feasible to a very limited sample of CCSNe, especially lacking those of SESNe. So-far the pre-explosion image is only available for two SNe Ib (iPTF 13bvn, \citealt{bersten14} and SN 2019yvr, \citealt{kilpatrick21}). Currently, modeling of the bolometric light curve is the main tool to infer the properties of the progenitor and the explosion, and most of them are based on the model established by \citet{arnett82}. For the hydrogen-poor SNe, the ejecta is mainly powered by the decay of the radioactive $^{56}$Ni/Co, and the diffusion time scale of the photon generated by the decay chain determines the width of the light curve. With the photospheric velocity compiled from the spectra at maximum light, the ejecta mass, the kinetic energy, and their mutual relation can be determined. However, previous research based on this method is mainly conducted at the photospheric phase, i.e., the period during which the ejecta is still optically thick, and the analyses constrain the nature of the outermost region of the ejecta. The interpretation regarding the global properties of the ejecta thus relies on extrapolation of the ejecta properties inward. 

In this work, we propose a complementary method to constrain the relation between the progenitor mass and the kinetic energy of SESNe, based on the observation at the nebular phase, i.e., several months after the explosion when the ejecta becomes transparent. \citet{fang22} reported a correlation between the [O~I]/[Ca~II] ratio, which serves as an indicator of the progenitor mass (\citealt{fransson89,jerkstrand15, kuncarayakti15, jerkstrand17,fang18,fang19,dessart21,fang22}), and the [O~I] width, which measures the characteristic expansion velocity of the oxygen-rich material (\citealt{taubenberger09,maurer10,fang22}), using a large sample of nebular spectra of 103 SESNe. In contrast to the observation at the photospheric phase, the nebular phase observation is sensitive to the properties in the dense innermost region where the explosion is initialized, and thus the explosion mechanism.

To build up the connection between the progenitor CO core mass ($M_{\rm CO}$) and the kinetic energy ($E_{\rm K}$) from theoretical aspect, we explode the helium star and CO star models generated by the one-dimensional stellar evolution code, Modules for Experiments in Stellar Astrophysics ($\texttt{MESA}$, \citealt{paxton11, paxton13, paxton15, paxton18, paxton19}), with a wide range of kinetic energy injected, using the SuperNova Explosion Code (\texttt{SNEC}, \citealt{snec15}). Omitting detailed spectrum synthesis calculations which would require massive computations, including a detailed treatment of the non-local thermal equilibrium (non LTE), we focus on the scaling relations between the model and the observed quantities. We especially apply the relation between the [O~I]/[Ca~II] ratio and the oxygen mass $M_{\rm O}$ based on the specific models by \citet{jerkstrand15}. The properly-weighted velocity is linked to the observed line width. The $M_{\rm CO}$-$E_{\rm K}$ relation is then established by linking the models to the [O~I]/[Ca~II]-[O~I] width correlation.

Finally, the $M_{\rm CO}$-$E_{\rm K}$ relation established from the nebular phase observation is compared to those derived from the early phase observation. The early phase and late phase observations are indeed probing different regions of the ejecta. The combined analysis of the observations from these two periods thus provides us a unique chance to scan through the ejecta from the outermost layer to the innermost region, which will be useful to reconstruct the full ejecta structure. Further, any possible inconsistency between the analyses based on the early phase and nebular phase observations will help to clarify what is still missing in the current assumptions on the ejecta structure, and improve our understanding of the ejecta dynamics.

The paper is organized as follows. In \S 2, we describe the numerical approaches, including the generation of the progenitor models, the mixing scheme, and the set up of the explosion. In \S 3, we introduce how the model quantities are connected to the observables, and derive the quantitative $M_{\rm CO}$-$E_{\rm K}$ relation based on the [O~I]/[Ca~II]-[O~I] width correlation. The possible affecting factors, including the dependence of [O~I]/[Ca~II] on $E_{\rm K}$ and the degree of microscopic mixing, and the effect of macroscopic mixing on the line width, are discussed in \S 4. The $M_{\rm CO}$-$E_{\rm K}$ relation established in this work is compared with the ones derived from the early phase observation in \S 5. The paper is closed with a summary of our findings in \S 6.

\section{Numerical approaches}
\subsection{{\rm \texttt{MESA}}: from pre-main-sequence to core-collapse}
The SN progenitor models are calculated using the one-dimensional stellar evolution code, Modules for Experiments in Stellar Astrophysics ($\texttt{MESA}$, \citealt{paxton11, paxton13, paxton15, paxton18, paxton19}). We start with $\texttt{MESA}$ version 11701 test suites, and closely follow the setup of $\texttt{example\_make\_pre\_ccsn}$ with minor modification. The inlists in this test suite include all processes involved in massive star evolution, including pre-main-sequence evolution, gradual burning of elements, removal of the outermost layers and the hydrodynamics of the iron-core infall. The zero-age-main-sequence masses ($M_{\rm{ZAMS}}$) are 13, 15, 18, 20, 23, 25, and 28 $M_{\odot}$. For all the models, we fix the metallicity ($Z$=0.02, i.e., the solar metallicity) and mixing length ($\alpha_{\rm env}$=3.0 in the hydrogen-rich envelop and 1.5 in the other regions). No rotation is introduced.

This paper mainly focuses on the pre-SN structure of a helium star (the progenitor of SNe IIb/Ib, if the hydrogen skin of SNe IIb is neglected) and a bare CO core (the progenitor of SNe Ic/Ic-BL), therefore the hydrogen envelope or the helium-rich layer should be removed before the explosion. There are several channels that may be responsible for the envelope-stripping process, i.e., binary mass transfer, stellar wind, or a combination of both (\citealt{heger03, sana12, groh13, smith14, yoon15, fang19}). However, after the helium burning is finished, the core structure will not be significantly affected by the presence or the absence of the outermost layers, therefore the detailed mass-loss mechanism is not important for the purpose of this work. The hydrogen envelope or the helium-rich layer is thus removed manually. We first evolve the star to the helium ignition phase without mass loss. After the helium in the center is exhausted, the mass loss rate is manually changed to 10$^{-3}$ $M_{\odot}$ yr$^{-1}$ (or 10$^{-4}$ $M_{\odot}$ yr$^{-1}$) for the removal of the hydrogen envelope (or the helium-rich layer), until the hydrogen (or helium) mass drops below 0.01 $M_{\odot}$ (or 0.12 $M_{\odot}$). Observationally, it has been indicated that SNe Ic/Ic-BL have a larger progenitor CO core mass than SNe IIb/Ib (\citealt{fang19,fang22,sun23}), therefore the helium-rich layer is stripped only for models with $M_{\rm ZAMS}$ not less than 18 $M_{\odot}$.
After the outer layers are removed, we calculate the subsequent evolution without mass loss until the Fe-core collapse. The inlists used to generate the progenitor models in this work are available on Zenodo under an open-source 
Creative Commons Attribution 4.0 International license: 
\dataset[doi:10.5281/zenodo.7740506]{https://zenodo.org/record/7740506}.

\begin{figure}[!htb]
\epsscale{1.05}
\plotone{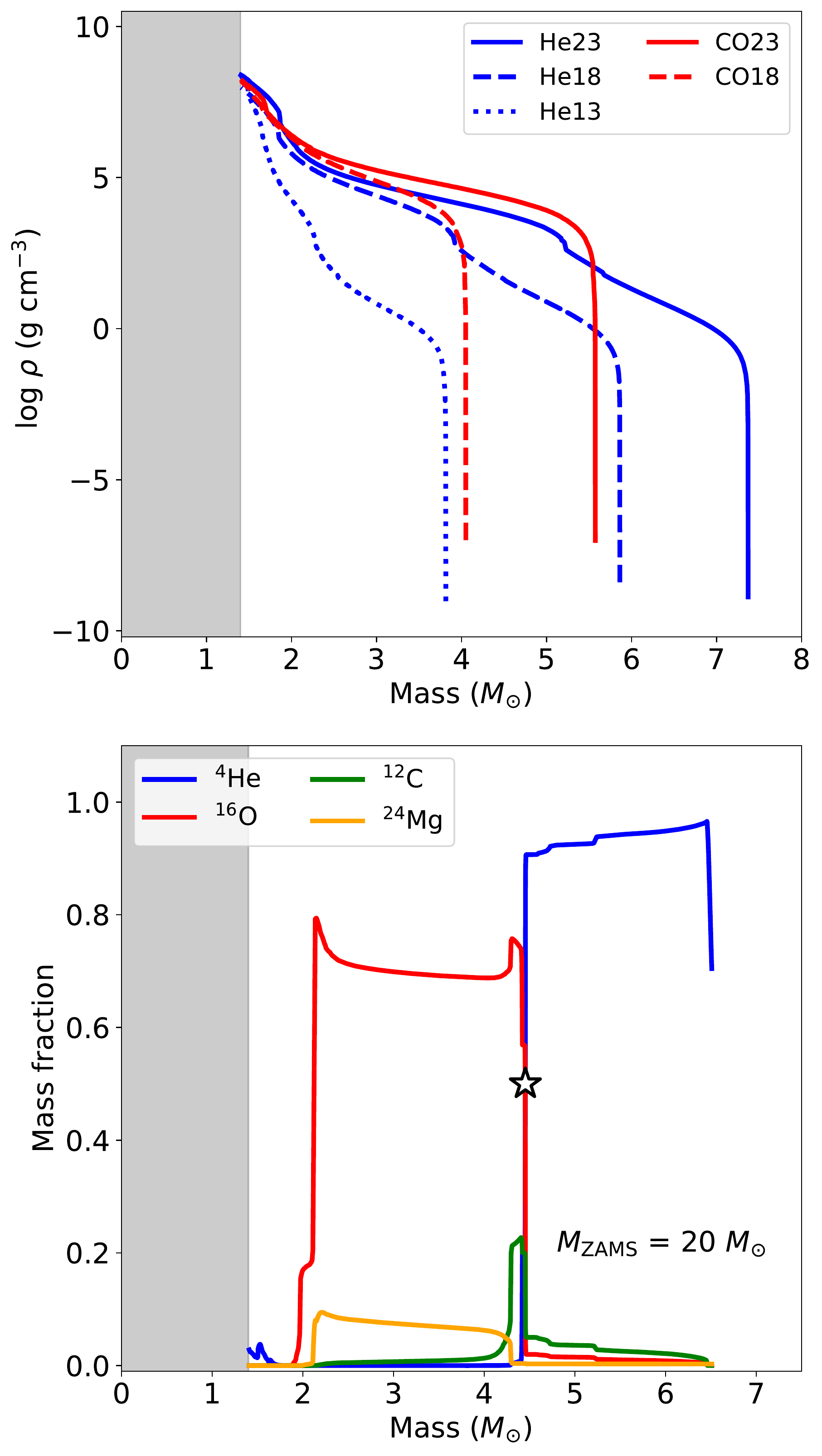}
\centering
\caption{$Upper~panel$: the density structures of the He stars with $M_{\rm ZAMS}$ = 13, 18, 23 $M_{\odot}$, and the bare CO stars with $M_{\rm ZAMS}$ = 18, 23 $M_{\odot}$; $Lower~panel$: the mass fractions of $^{4}$He, $^{12}$C, $^{16}$O and $^{24}$Mg for the helium star with $M_{\rm ZAMS}$ = 20 $M_{\odot}$. The star marks the mass coordinate of the CO core boundary. The shaded regions in the two panels represent the region collapsing into the compact remnant.}
\label{fig:pre_SN_pro}
\vspace{4mm}
\end{figure}

In the upper panel of Figure \ref{fig:pre_SN_pro}, we show the pre-SN density structures of the helium stars with $M_{\rm ZAMS}$ = 13, 18, 23$M_{\odot}$, and the bare CO core with $M_{\rm ZAMS}$ = 18, 23$M_{\odot}$. The mass fractions of $^{4}$He, $^{12}$C, $^{16}$O and $^{24}$Mg for the helium star with $M_{\rm ZAMS}$ = 20$M_{\odot}$ is also plotted in the lower panel of Figure \ref{fig:pre_SN_pro} for illustration.

\begin{figure}[!htb]
\epsscale{1.05}
\plotone{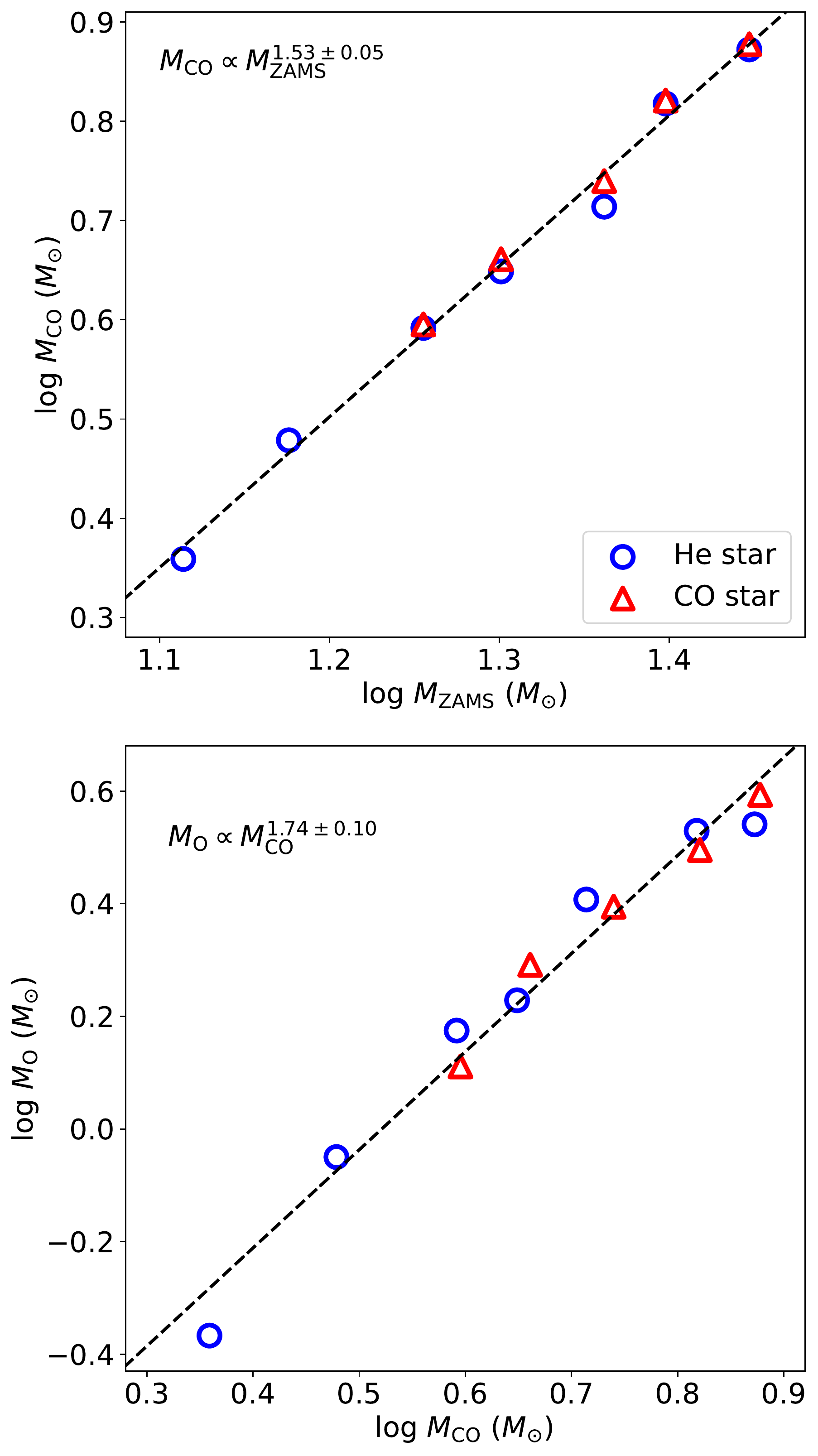}
\centering
\caption{$Upper~panel$: The relation between the CO core mass and the $M_{\rm ZAMS}$ of the progenitor models; $Lower~panel$: The relation between the oxygen mass and the CO core mass.}
\label{fig:pre_SN_corr}
\vspace{4mm}
\end{figure}

Some properties of the progenitor models are summarized in Table \ref{tab:progenitor}. In this work, the outer boundary of the CO core is defined to be the mass coordinate with the helium mass fraction $X_{\rm He}$ = 0.5 (as marked by the black star in the lower panel of Figure \ref{fig:pre_SN_pro}); the CO core mass ($M_{\rm CO}$) refers to the mass coordinate at the CO core outer boundary. The mass of the oxygen is 

\begin{equation}
   M_{\rm O} ~ = ~ \sum X_{\rm O}(m_{\rm i})\Delta m_{\rm i},
\end{equation}
where $X_{\rm O}(m_{\rm i})$ is the oxygen mass fraction of the grid centered at $m_{\rm i}$. The CO core mass ($M_{\rm CO}$) is strongly correlated with $M_{\rm ZAMS}$. The linear regression (in logrithium scale) gives
\begin{equation}
   M_{\rm CO} ~ \propto ~ M_{\rm ZAMS}^{1.53\pm 0.05}.
\end{equation}
The oxygen mass $M_{\rm O}$ is also correlated with $M_{\rm CO}$, and scales as 
\begin{equation}
   M_{\rm O} ~ \propto ~ M_{\rm CO}^{1.74\pm 0.10}.
\label{eq:O_CO_preSN_relation}
\end{equation}
The above correlations are plotted in Figure \ref{fig:pre_SN_corr}. The effect of the attached helium-rich layer on the CO core properties is negligible.

In the following, we use the term HeXX (or COXX) to represent helium star (or bare CO star) model with $M_{\rm ZAMS}$ = XX $M_{\odot}$. For example, He15 and CO20 represent a helium star with $M_{\rm ZAMS}$ = 15$M_{\odot}$ and a bare CO star with $M_{\rm ZAMS}$ = 20$M_{\odot}$, respectively.

\begin{deluxetable}{cccccc}[t]
\centering
\label{tab:progenitor}
\tablehead{
\colhead{Model} & \colhead{$M_{\rm ZAMS}$} & \colhead{$M_{\rm pre-SN}$}&\colhead{$M_{\rm CO}$}&\colhead{$M_{\rm env}$}&\colhead{$M_{\rm O}$}
}
\startdata
He13&13&3.82&2.27&1.55&0.43\\
He15&15&4.74&2.99&1.75&0.89\\
He18&18&5.86&3.90&1.96&1.50\\
He20&20&6.51&4.45&2.06&1.70\\
He23&23&7.37&5.18&2.20&2.56\\
He25&25&8.88&6.57&2.31&3.37\\
He28&28&9.96&7.45&2.51&3.47\\
CO18&18&4.05&3.94&-&1.29\\
CO20&20&4.67&4.58&-&1.96\\
CO23&23&5.57&5.49&-&2.48\\
CO25&25&6.74&6.62&-&3.13\\
CO28&28&7.67&7.54&-&3.92\\
\enddata
\caption{Summary of the pre-SN model properties in this study. The helium star models (prefix 'He') are listed at the upper part and the CO star models (prefix 'CO') are at the lower part. For each model, we give the ZAMS mass ($M_{\rm ZAMS}$), the mass at core collapse ($M_{\rm pre-SN}$), the CO core mass defined by $X_{\rm He}$=0.5 ($M_{\rm CO}$), the mass of the helium-rich layer ($M_{\rm env}$) and the total oxygen mass in the ejecta ($M_{\rm O}$). These quantities are all in the unit of solar mass $M_{\odot}$.}
\end{deluxetable}

\subsection{$^{56}$Ni mixing}
During the shock wave propagation, Rayleigh-Taylor and Richtmyer-Meshkov instabilities will develop, resulting in effective mixing of the ejecta (\citealt{mixing03, mixing06, mixing15}). Such instabilities are important to the dynamics of the ejecta, but cannot be accurately modeled by 1D simulations. 
The effect of large scale material mixing in the ejecta of CCSNe, with a focus on the radioactive energy source $^{56}$Ni, have long been studied (\citealt{ensman88, shigeyama90a, shigeyama90b, woosley95, sauer06, dessart11, dessart12, dessart15, dessart16, bersten13, piro13, cano14, yoon19, moriya20, teffs20}). However, the degree of mixing in the CCSN ejecta, and its possible dependence on the SNe progenitor, are difficult to constrain from observation. Based on the light curves of a large sample of SESNe from the Carnegie Supernova Project (CSP; \citealt{hamuy06}), \cite{taddia18} concluded that SNe IIb/Ib show variation in the degree of $^{56}$Ni mixing, while for SNe Ic the $^{56}$Ni is fully mixed into the ejecta with few exceptions. By studying the color curve evolution of SESNe, \citet{yoon19} also suggest that $^{56}$Ni is only mildly mixed into the helium-rich layer of SNe IIb/Ib, while the ejecta of SNe Ic is fully mixed. This is further supported by the study on the evolution of photospheric velocity at very early phases. \citet{moriya20} calculate the photospheric velocity evolution of SESNe with different degrees of $^{56}$Ni mixing, and the models are further applied to the individual object, SN 2007Y. For this SN Ib, its photospheric velocity evolution matches well with the model where $^{56}$Ni is only mixed into about half of the ejecta in the mass coordinate.

\begin{figure}[!htb]
\epsscale{1.15}
\plotone{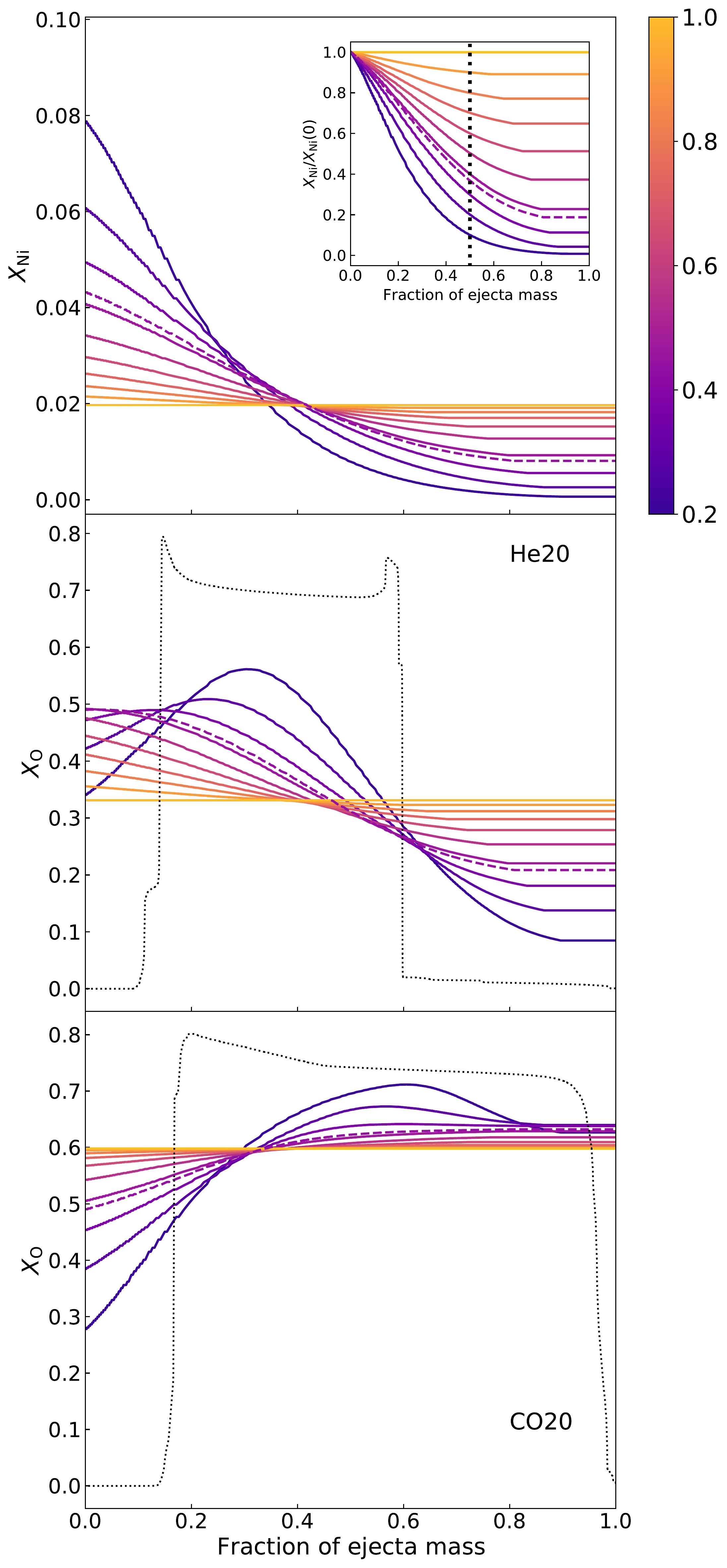}
\centering
\caption{$Upper~panel$: The $^{56}$Ni mass fraction of He20 model with different degrees of mixing, which is defined by Equation \ref{eq:f_define} and are labeled by different colors. The insert panel is the $^{56}$Ni fraction divided by its maximum. The black dashed line marks the mid-point of the ejecta; $Middle~panel$: The $^{16}$O mass fraction of He20 model with different degrees of mixing. The $^{16}$O mass fraction of the pre-SN model (un-mixed) is shown by the black dotted line for comparison; $Lower~panel$: The $^{16}$O mass fraction of CO20 model with different degrees of mixing. The $^{16}$O mass fraction of the pre-SN model is shown by the black dotted line for comparison.}
\label{fig:ni_mix}
\end{figure}

$^{56}$Ni is the explosive-burning product, and its distribution is not strongly constrained from the current models. In this work, $^{56}$Ni is phenomenologically mixed with the method introduced as follow. First, 0.1 $M_{\odot}$ of $^{56}$Ni is uniformly put in the innermost 10\% (in mass coordinate) of the ejecta by hand. The ejecta is then artificially mixed by the "boxcar" averaging (\citealt{kasen09, dessart12, dessart13, snec15})\footnote{The readers may refer to the notes of $\texttt{SNEC}$ for the details of this procedure.}. We define 
\begin{equation}
        f= \frac{X_{\rm Ni}(M_{\rm r} = 0.5M_{\rm ejecta})}{X_{\rm Ni}(M_{\rm r} = 0)},
\label{eq:f_define}
\end{equation}
i.e., the ratio of the $^{56}$Ni fraction ($X_{\rm Ni}$) at the mid-point of the ejecta and that at the center of the ejecta. Here $M_{\rm r}$ is the mass coordinate after the remnant is excised. In this work, this ratio is employed to characterize the mixing degree of the ejecta. For each progenitor model, the degree of mixing $f$ is varied from 0.1 to 1.0 ("fully mixed") with 0.1 increment by adjusting the width of the boxcar, as shown in the upper panel of Figure \ref{fig:ni_mix}. The other elements in the ejecta are accordingly mixed by the boxcar averaging scheme. The $^{16}$O distributions of the mixed ejecta with different $f$ values are shown in the middle (He20 model) and lower panels (CO20 model) in Figure \ref{fig:ni_mix}.

\citet{yoon19} characterized the $^{56}$Ni distribution by

\begin{equation}
        X_{\rm Ni}(M_{\rm r}) \propto {\rm exp}\left(-\left[\frac{M_{\rm r} ¥}{f_{\rm Y19}M_{\rm ejecta}}\right]^2\right).
\end{equation}
By studying the early-phase color curve evolution of a sample of helium-rich SNe, \citet{yoon19} conclude that for these objects, $^{56}$Ni is only mixed up to the mid-point of the helium-rich envelope, or $f_{\rm Y19}$= 0.3 to 0.5, which corresponds to $f= 0.368$ in the present work.
Therefore, in the following analysis, we employ $f$=0.368 as the default case, unless explicitly mentioned.
The effect of large scale mixing is discussed in \S 4. 

\subsection{{\rm \texttt{SNEC}}: explosion hydrodynamics}
Once the progenitor models have evolved to the time of core collapse, they are used as the input models of the hydrodynamics simulation of a supernova explosion. In this work, we use the SuperNova Explosion Code (\texttt{SNEC}, \citealt{snec15}) to solve the hydrodynamic evolution of the SN ejecta. 

Before the set up of the explosion, the materials are firstly mixed manually as introduced above. The explosion is initiated as the "thermal bomb" mode, i.e., the explosion energy is initially injected into a small mass range (${\rm \Delta}M$=0.1$M_{\odot}$) and the injection lasts for 0.2 seconds. We vary the final energies (thermal energies plus kinetic energies) of the explosions ($E_{\rm K}$) from $\sim 10^{51}$ erg to $10 \times 10^{51}$ erg with 0.5$\times 10^{51}$ erg increments. In the following, we refer $10^{51}$ erg as 1 foe. The inner 1.4$M_{\odot}$ is excised to account for the compact remnant formation. 

\begin{figure}[!htb]
\epsscale{1.05}
\plotone{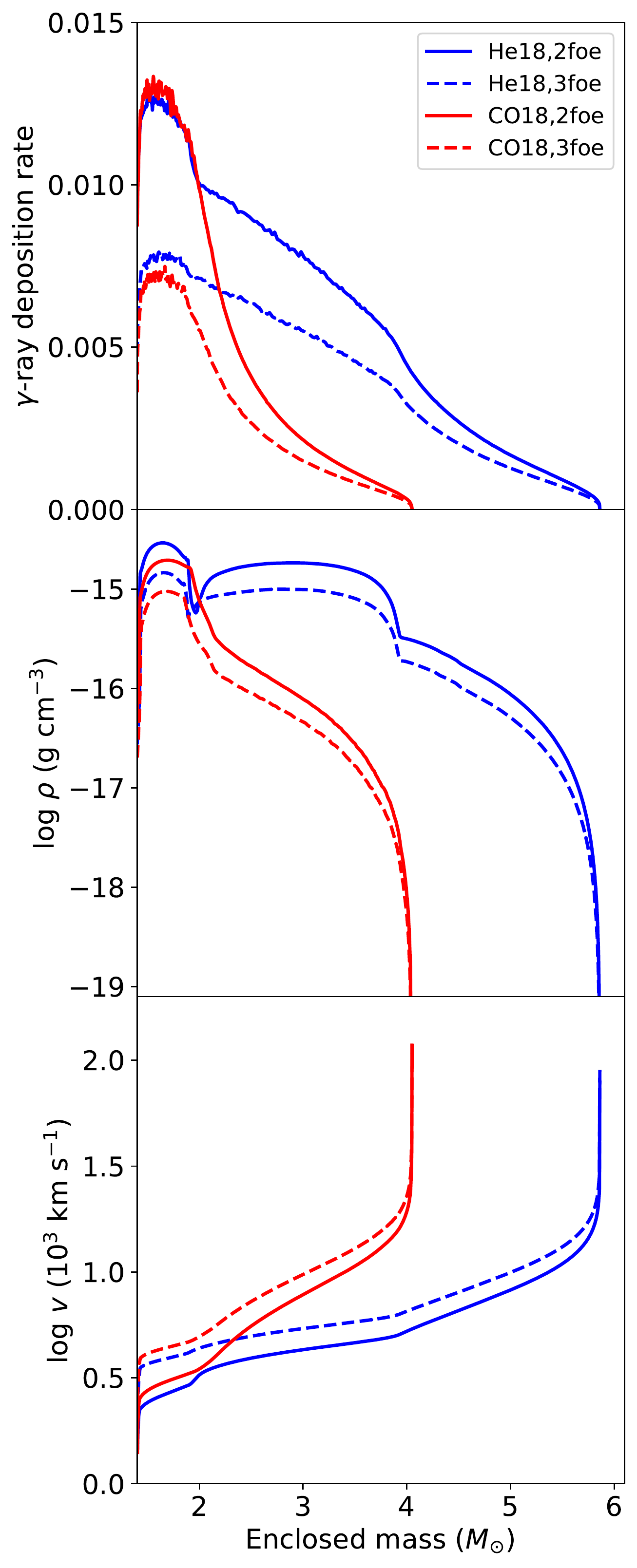}
\centering
\caption{The physical properties of the ejecta of He18 and CO18 models (labeled by different colors) with different kinetic energies (labeled by different line styles). $Upper~panel$: The $\gamma$-ray deposition rate; $Middle~panel$: The density profile; $Lower~panel$: The velocity profile. These properties are shown for 220 days after the explosion.}
\label{fig:snec_pro}
\end{figure}

The $\gamma$-ray deposition rates, as well as density and velocity profiles of the ejecta ($t$=220 days after the explosion) of He18 and CO18 models with different kinetic energies, are plotted in Figure \ref{fig:snec_pro}.

\section{Connecting models to observables}

\subsection{Oxygen mass and [O~I]/[Ca~II] }
The [O~I]/[Ca~II] ratio is frequently employed as an indicator for the CO core mass of the progenitor. The oxygen mass is mainly determined by the progenitor CO core mass, to which the Ca mass is insensitive. However, the dependence of the [O~I]/[Ca~II] ratio on the O mass of the progenitor has not been quantitatively clarified from observation. 

\begin{figure}[!htb]
\epsscale{1.05}
\plotone{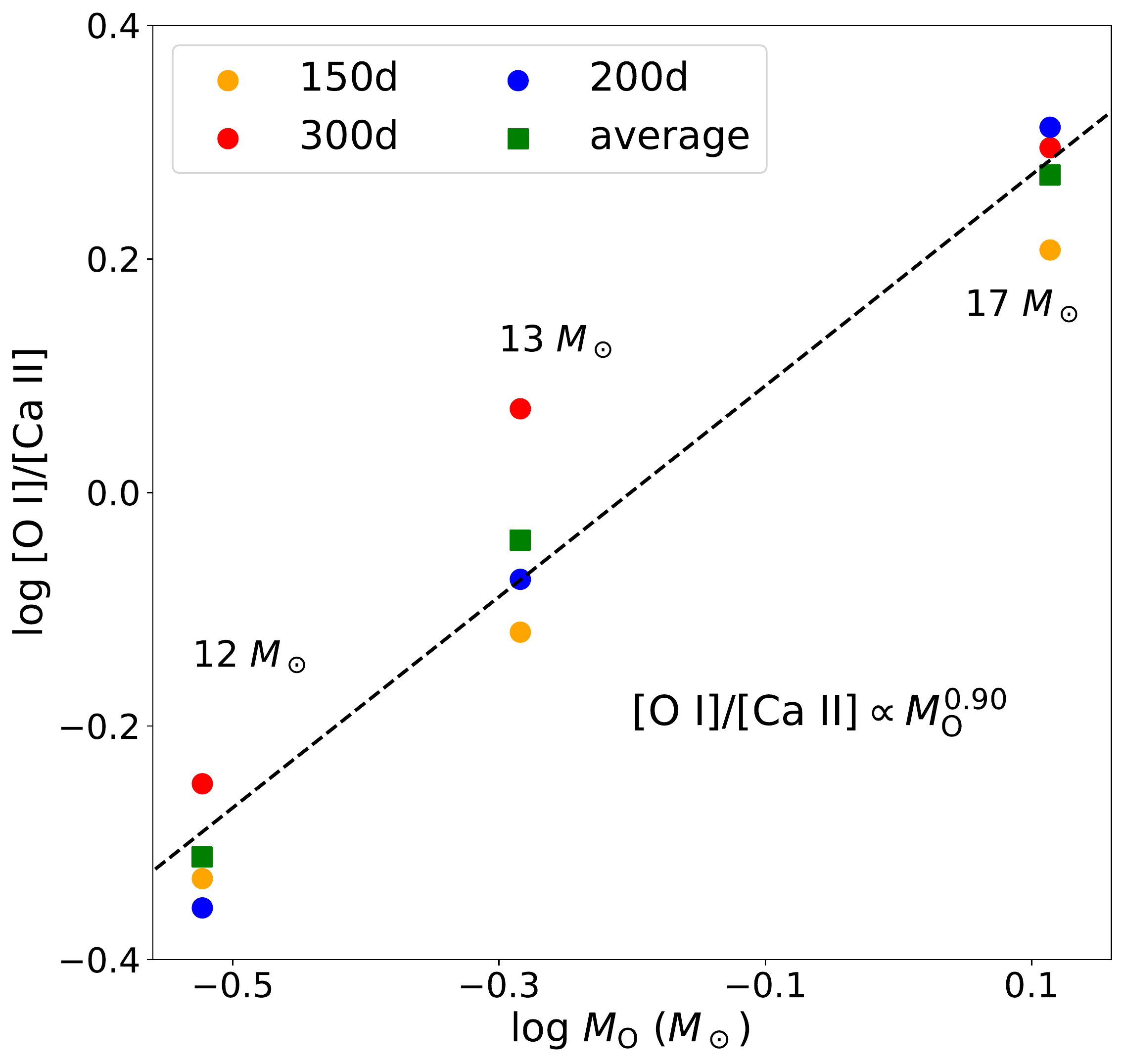}
\centering
\caption{The relation between the [O~I]/[Ca~II] of SNe IIb model spectra (\citealt{jerkstrand15}) and the O mass of the ejecta. Measurements at different phases are labeled by different colors.}
\label{fig:model_ratio}
\end{figure}

The SNe IIb spectral models of \citet{jerkstrand15} provide a possible constraint on the connection between the [O~I]/[Ca~II] ratio and the O mass of the ejecta. We apply the same method as \citet{fang22} to the model spectra to measure the [O~I]/[Ca~II] ratios, which are then compared with the O mass of the progenitor models in \citet{jerkstrand15}, as shown in Figure \ref{fig:model_ratio}. The average phase of the nebular SESNe in the sample of \citet{fang22} is 220$\pm$80 days, therefore the measurement is limited to the model spectra at 150, 200 and 300 days. If we assume [O~I]/[Ca~II]$\propto M_{\rm O}^{\alpha}$, the linear regression in logarithmic scale gives $\alpha$ = 0.82 (300 days) and 1.01 (200 days). On average, we have
\begin{equation}
    {\rm [O~I]/[Ca~II]}~\propto~M_{\rm O}^{0.90\pm 0.09}.
\label{eq:Mo_ratio_relation}
\end{equation}
This relation will be applied to connect the [O~I]/[Ca~II] and the $M_{\rm O}$ of the helium star models in this work. Lacking consistent nebular model spectra of SNe Ic, whether this relation can be applied to the bare CO star models remains uncertain. While keeping this caveat in mind, it is illustrative to extend this relation to the helium-deficient models to compare with the observed SNe Ic/Ic-BL.

It should be noted that [O~I]/[Ca~II] is not only determined by the oxygen mass $M_{\rm O}$, but also affected by the physical properties of the ejecta, including temperature, density, and so on. Here we have assumed that these quantities are ultimately determined only by the progenitor mass, therefore their effects on [O~I]/[Ca~II] are absorbed in the scaling index of $M_{\rm O}$.  Discussion on the variation form of Equation \ref{eq:Mo_ratio_relation} is left to \S 4.1. We further note that we have fixed the metallicity in this investigation (assuming the solar metallicity). The metallicity will introduce a mass-independent factor to a problem, but the observed variation of the metallicity at the SN site is not exceedingly large (see for example \citealt{modjaz08}), therefore its effect on the bulk statistics should be negligible.

\subsection{Ejecta velocity and [O~I] width}
The SN ejecta is powered by the deposited $\gamma$-rays originally emitted from the decays of $^{56}$Ni and $^{56}$Co, and the heating process is balanced by the line emissions of the elements in each shell. In the expanding ejecta, each mass shell has different expansion velocity, therefore the centers of the emission lines are Doppler shifted. In SNe, the Doppler effect is the dominating broadening factor of the lines, therefore the widths of the emission lines can inversely be utilized to determine the velocity scales of the corresponding emitting elements. 

Following the explosion of a massive star, the ejecta expands homologously with $V(r,t)=r/t$, where $V(r,t)$ is the expansion velocity of the mass shell located at radius $r$ at time $t$. In the spherically symmetric case, the specific flux at frequency $\nu$ is
\begin{equation}
    F_{\nu}\propto\int_{V(\nu)}^{V_{\rm max}}j(V)VdV.
\label{eq:line_profile}
\end{equation}
Here, $V_{\rm max}$ is the outermost velocity of the ejecta and $V({\nu})$ = $\frac{\nu_0-\nu}{\nu_0}c$, where $\nu_0$ is the rest frame frequency of the emission and $c$ is the light speed. The emission coefficient in the mass shell with expansion velocity $V$ is $j(V)$. The readers may refer to \citet{jerkstrand17} for the detailed discussion on the formation of the nebular line profile.

\begin{figure}[!htb]
\epsscale{1.05}
\plotone{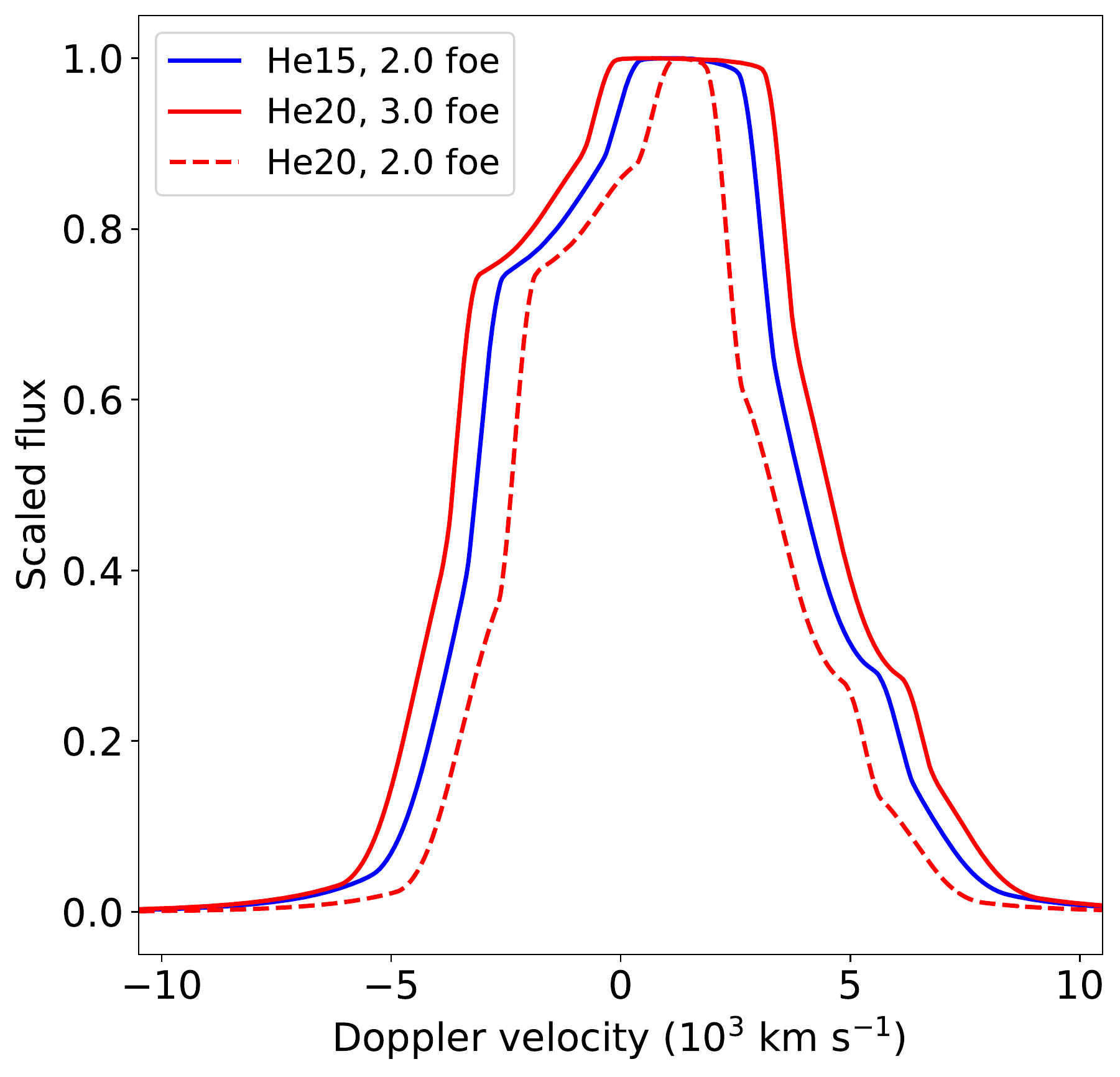}
\centering
\caption{The line profile constructed by Equations \ref{eq:line_profile} and \ref{eq:emissivity} for He15 and He20 models (labeled by different colors) with different kinetic energies (labeled by different line styles).}
\label{fig:line_profile_example}
\end{figure}

The rate of radioactive energy deposited in a mass shell is $\epsilon_{\rm rad}d$ by definition, where $d$ is the local $\gamma$-ray deposition function per mass. Here, $\epsilon_{\rm rad}$ is the rate of energy release per gram of radioactive nickel. We assume that the deposited energy is re-emitted by [O~I] at a rate of $X_{\rm O ~I} \epsilon_{\rm rad}d$ (see below), where $X_{\rm O ~I}$ is the mass fraction of neutral oxygen. Therefore, we have
\begin{equation}
    j_{\rm [O~I]} \propto \rho X_{\rm O ~I} \epsilon_{\rm rad}d.
\label{eq:emissivity}
\end{equation}
By assuming $X_{\rm O ~I}$ $\sim$ $X_{\rm O}$ and $L(6300)/L(6363)$=3 (optical thin limit), the [O~I] profile can be constructed by Equation \ref{eq:line_profile}. Some examples are illustrated in Figure \ref{fig:line_profile_example}.

Indeed, when the oxygen dominates the cooling, its mass fraction would not sensitively affect the line strength (i.e., the temperature is anyway determined to balance the heating and cooling rates). However, we introduce a factor $X_{\rm O}$ here, to account for the mixing effect as introduced above, since $X_{\rm O}$ traces the fraction of the O-rich material in a given volume once it is macroscopically mixed with other characteristic nuclear-burning layers. We note that we are not concerned with the absolute flux scale, and therefore this procedure is justified as long as $X_{\rm O}$ in the original (unmixed) ejecta is roughly constant within the O-rich region (which is indeed the case; Figure \ref{fig:pre_SN_pro}).

\begin{figure}[hbt]
\epsscale{1.05}
\plotone{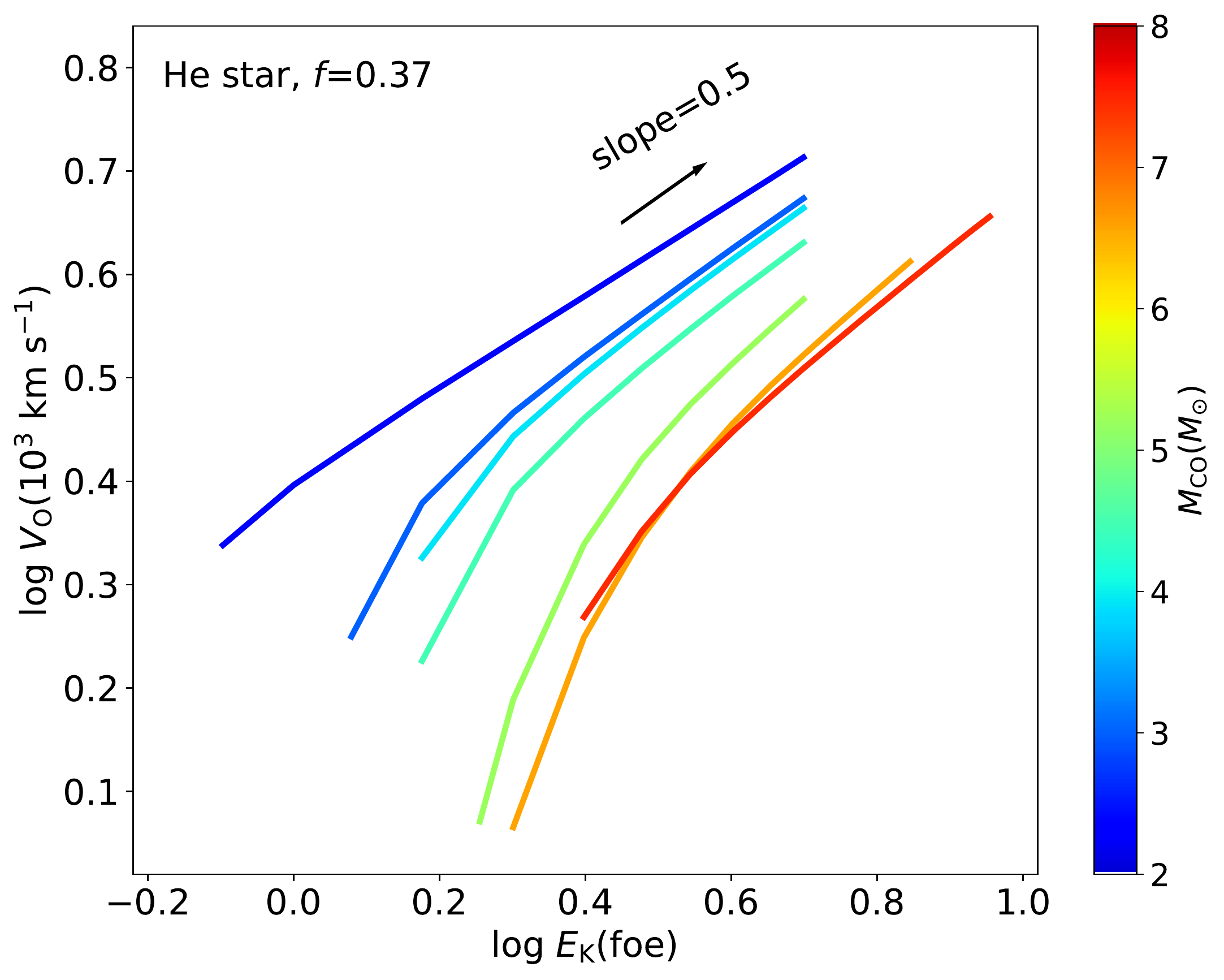}
\centering
\caption{The relation between $V_{\rm O}$ and the $E_{\rm K}$ of the ejecta. The mixing degree $f$ is fixed to be 0.368 to match with the results in \cite{yoon19}. The colorbar indicates $M_{\rm CO}$ of the progenitor. For a fixed $M_{\rm CO}$ (or progenitor model), the slope is very close to 0.5 in logarithmic scale at the relatively high $E_{\rm K}$ end.}
\label{fig:snec_velo}
\end{figure}

We apply the same line width measurement method as \citet{fang22} to the model spectra, i.e., half of the wavelength range (or velocity range) that contains 68\% of the total emission flux of the model [O~I] profile. The measured line width is dependent on both $M_{\rm O}$ and $E_{\rm K}$. As shown in Figure \ref{fig:snec_velo}, for a fixed He star model (therefore fixed $M_{\rm O}$), the measured width $V_{\rm O}$ scales as $V_{\rm O}\propto E_{\rm K}^{0.5}$.

\subsection{[O~I]/[Ca~II]-[O~I] width correlation}
In \citet{fang22}, a correlation between the [O~I]/[Ca~II] and [O~I] width is discerned, based on a large sample of SESN nebular spectra ($N$=103). For the helium-rich SNe IIb/Ib, the correlation is significant, while it is not clearly discerned for the helium-deficient SNe Ic/Ic-BL. The correlation itself, along with its different dependence on the SNe sub types, can be qualitatively explained if the kinetic energy of the explosion is correlated with the progenitor CO core mass. In this work, we will derive the quantitative relation between the CO core mass and the kinetic energy $E_{\rm K}$ that is required to reproduce the correlation.

First the observed line width ${\Delta \lambda}$ is transformed to the typical velocity scale $V_{\rm Obs}$ by
\begin{equation}
    V_{\rm Obs}=\frac{\Delta \lambda}{6300~{\rm \AA}}\times c.
\end{equation}
To connect the progenitor models to the observables, we assume [O~I]/[Ca~II]$\propto M_{\rm O}^{0.90}$ (see \S 3.1). The oxygen mass $M_{\rm O}$ and the measured [O~I] width $V_{\rm O}$ of the models are multiplied by constants to match the He13 model with $E_{\rm K}$=0.94 foe \citep[see][]{fremling16} with the [O~I]/[Ca~II] and $V_{\rm Obs}$ values of iPTF 13bvn. These calibrations give

\begin{equation}
    {\rm log}\frac{\rm [O ~I]}{\rm [Ca ~II]}=0.9\times{\rm log}\frac{M_{\rm O}}{M_{\odot}}+0.03,\\
\label{eq:model_to_obs1}
\end{equation}
and
\begin{equation}
    {\rm log}\frac{V_{\rm Obs}}{\rm 10^3~ km~s^{-1}}={\rm log}\frac{V_{\rm O}}{\rm 10^3 km~s^{-1}}-0.07.\\
\label{eq:model_to_obs2}
\end{equation}

\begin{figure}[hbt]
\epsscale{1.05}
\plotone{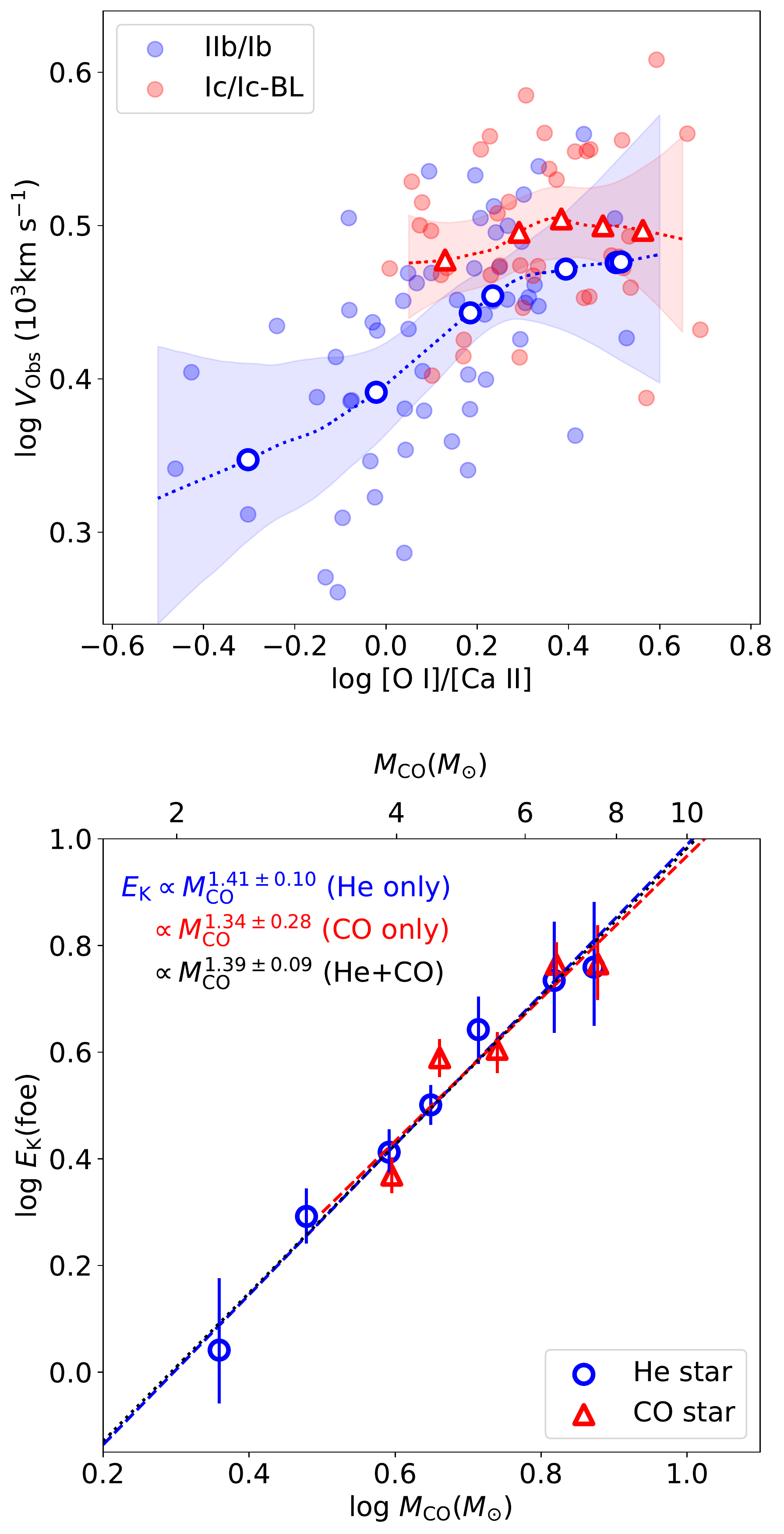}
\centering
\caption{$Upper~panel$: The observed [O~I]/[Ca~II]-[O~I] width correlation. The helium-rich SNe (type IIb + Ib) and the helium-deficient SNe (type Ic + Ic-BL) are labeled by different colors. The dashed lines are the local non-parametric regressions to the observation. The open-squares and open-triangles mark the helium star and CO star models; $Lower~panel$: The relation between the CO core mass $M_{\rm CO}$ of the models and the kinetic energy required to produce the observed [O~I]/[Ca~II]-[O~I] width correlation. The helium star and CO star models are labeled by different colors and markers. The dashed lines are the linear regressions to the He star models (blue), CO star models (red), He star + CO star models (black). The $M_{\rm CO}$-$E_{\rm K}$ relations of the helium-rich and helium-deficient SNe are almost identical.}
\label{fig:main}
\end{figure}

The upper panel of Figure \ref{fig:main} is the observational result of \citet{fang22}. The local non-parametric regression is performed to the SNe IIb/Ib and SNe Ic/Ic-BL respectively, as marked by the dashed lines. The shaded regions are the 95\% confidence intervals (CI).  For a specific model, its $M_{\rm O}$ is transformed to the observed [O~I]/[Ca~II] using Equation \ref{eq:model_to_obs1}. With the results from the local non-parametric regression, we derive $V_{\rm Obs}$ required for this progenitor model to reproduce the observed correlation, as marked in the upper panel of Figure \ref{fig:main}, which is then further transformed to the model velocity ($V_{\rm O}$) using Equation \ref{eq:model_to_obs2}. The velocity, $V_{\rm O}$, is transformed to the kinetic energy of the specific model using the relations in Figure \ref{fig:snec_velo}. The result is summarized in Table \ref{tab:kinetic_energy}.

It is clear that the kinetic energy of the explosion is required to be larger for He star model with a larger amount of oxygen (therefore larger $M_{\rm ZAMS}$) to produce the observed [O~I]/[Ca~II]-[O~I] width correlation. This is already pointed out by the qualitative analysis of \citet{fang22}. The relation between the CO mass ($M_{\rm CO}$) and kinetic energy ($E_{\rm K}$) is shown in the lower panel of Figure~\ref{fig:main}. If only the He star models are included, we have the scaling relation
\begin{equation}
    E_{\rm K}\propto M_{\rm CO}^{1.41\pm 0.10}.
\label{eq:Ek_O_relation_He_only}
\end{equation}
If Equation \ref{eq:model_to_obs1} is also applied to the CO core models, with the similar practice, we derive the relation between the $M_{\rm CO}$ and $E_{\rm K}$ for these models, which is also plotted in the lower panel of Figure ~\ref{fig:main}.
For the CO core models, the scaling relation is
\begin{equation}
    E_{\rm K}\propto M_{\rm CO}^{1.34\pm 0.28},
\label{eq:Ek_O_relation_CO_only}
\end{equation}

If the He star and the CO core models are combined, the relation between $M_{\rm CO}$ and $E_{\rm K}$ then becomes
\begin{equation}
    E_{\rm K}\propto M_{\rm CO}^{1.39\pm 0.09},
\label{eq:Ek_O_relation_He_plus_CO}
\end{equation}
which is similar to Equation \ref{eq:Ek_O_relation_He_only} where only helium stars are included. The correlation is significant with Spearman's rank coefficient $\rho$=0.98 and $p\textless$0.0001. This implies the kinetic energy of SNe Ic has the same dependence on $M_{\rm CO}$ (or $M_{\rm ZAMS}$) as their helium-rich counterparts, and possibly SNe IIb/Ib and SNe Ic share the same explosion mechanism despite the different degrees of the helium-rich layer stripping. 

It should be noted that the scaling relation between the [O~I]/[Ca~II] ratio and $M_{\rm O}$ (Equation \ref{eq:Mo_ratio_relation}) is empirically derived from the nebular helium-rich SNe models of \citet{jerkstrand15}, therefore it is not necessarily valid for the helium-deficient SNe. The application of this relation to the CO core models and SNe Ic/Ic-BL is only for illustrative purpose. Further discussion on this topic is left to \S 4.1.

\begin{deluxetable}{cccc}[t]
\centering
\label{tab:kinetic_energy}
\tablehead{
\colhead{$E_{\rm K}$(foe)} & Lower& Middle &Upper
}
\startdata
He13&0.88&1.10&1.49\\
He15&1.75&1.96&2.20\\
He18&2.38&2.58&2.84\\
He20&2.92&3.17&3.43\\
He23&3.81&4.39&5.03\\
He25&4.36&5.42&6.95\\
He28&4.49&5.75&7.57\\
\hline
CO18&2.18&2.34&2.51\\
CO20&3.60&3.88&4.19\\
CO23&3.66&4.03&4.31\\
CO25&5.36&5.82&6.36\\
CO28&5.02&5.83&6.84\\
\enddata
\caption{The kinetic energy required to reproduce the observed correlation for the progenitor models. The upper and lower limits are transformed from the 95\% CI.}
\end{deluxetable}

\section{Discussion}
\subsection{Scaling relation}
\subsubsection{Factors that might affect [O~I]/[Ca~II]}
In the previous sections, we have assumed that [O~I]/[Ca~II] is determined only by the oxygen mass $M_{\rm O}$, which is based on the assumption that other affecting factors (density, temperature, etc.) are also dependent on the progenitor mass so that their effects on [O~I]/[Ca~II] are all absorbed into the dependence on $M_{\rm O}$. However, this assumption is not necessarily valid. The calcium emission [Ca~II] comes from the explosive-nucleosynthesis region, therefore its strength may well be affected by the kinetic energy of the explosion. Further, calcium is an efficient coolant. If a certain amount of calcium (mass fraction larger than 10$^{-3}$) is microscopically mixed into the oxygen-rich shell through diffusion, the strength of the [Ca~II] will dominate the [O~I] and the [O~I]/[Ca~II] ratio will be reduced (\citealt{fransson89,maeda07,dessart20}). These two factors, i.e., (1) the kinetic energy and (2) the microscopic mixing, will affect the [O~I]/[Ca~II] ratio as follows:


\begin{itemize}
    \item Kinetic energy: the kinetic energy will affect the [O~I]/[Ca~II] in two aspects: (1) The density of the ejecta. For the same pre-SN structure, the increase of the kinetic energy will increase the expansion velocity of the expelled material, resulting in low density ejecta. The assumption that the [O~I] and [Ca~II] dominate the emission from the O-rich shell and the explosive-nucleosynthesis region, respectively, is more valid when the density is lower. If the density of the O-rich shell increases, the contribution from Mg I] 4571 and [O~I] 5577 becomes non-negligible. For the explosive-nucleosynthesis region, the Ca II H\&K, NIR triplet and Si I 1.099 $\mu$m become strong when the density of this region increases. However, the emissions from the explosive-nucleosynthesis region is more sensitive to the density, therefore the decrease of the density (or increase of the explosion energy) will decrease the [O~I]/[Ca~II] ratio (\citealt{fransson89}); (2) nucleosynthesis: the amount of the newly synthesized elements, including calcium, generally increases with the explosion energy (\citealt{woosley02,limongi03}). The strength of the [Ca~II] thus traces the amount of the explosive-nucleosynthesis region. The increase of the explosion energy will therefore decrease the [O~I]/[Ca~II] ratio.
    \item Microscopic mixing: The [Ca~II] is mostly emitted by the newly synthesized calcium in the explosive burning ash (\citealt{jerkstrand15}). The microscopic mixing is not expected to occur during the explosion because the diffusion time scale is long, as inferred from the chemical inhomogeneity of Cas A (\citealt{ennis06}). However, if the pre-existing calcium, which is synthesized during the advanced stage of massive star evolution, is microscopically mixed into the O-rich shell before the explosion, its contribution to the [Ca~II] can become significant, and the [O~I]/[Ca~II] ratio will decrease because [Ca~II] is a more effective coolant than [O~I] (\citealt{dessart21}). The microscopic mixing may happen during the Si burning stage through the merger of the Si-rich and O-rich shell, and the occurrence rate is higher for a more massive progenitor between 16 to 26 $M_{\odot}$ (\citealt{collins18,dessart20}).
\end{itemize}
In conclusion, both the increase of the kinetic energy $E_{\rm K}$ and the diffusion of the calcium into the O-rich shell will tend to reduce the [O~I]/[Ca~II] ratio. 

In \S 3, the derivation of the $M_{\rm CO}$-$E_{\rm K}$ relation (Equation~\ref{eq:Ek_O_relation_He_plus_CO}) is based on the assumption that the [O~I]/[Ca~II] ratio is determined \emph{only} by the oxygen content of the progenitor (Equation~\ref{eq:model_to_obs1}). As stated above, this assumption is not necessarily valid. The relations between the [O~I]/[Ca~II] ratio and $E_{\rm K}$, as well as the microscopic mixing, are complicated, and would possibly affect the $M_{\rm CO}$-$E_{\rm K}$ relation. It is therefore important to examine how the $M_{\rm CO}$-$E_{\rm K}$ relation is altered if the above two factors are taken into consideration. However, a detailed treatment on this topic would require a large grid of stellar evolution models and radioactive transfer simulations, which is beyond the scope of this paper. In this section, we attempt to quantify the effects of these two factors on the $M_{\rm CO}$-$E_{\rm K}$ relation by including them into the scaling relation of [O~I]/[Ca~II] ratio and $M_{\rm O}$ in the power-law form. Equation \ref{eq:model_to_obs1} then becomes
\begin{equation}
    {\rm log} \frac{[{\rm O~I}]}{[{\rm Ca~II}]}=(0.90 - \alpha)\times{\rm log}M_{\rm O}-\beta\times{\rm log}E_{\rm K},
\label{eq:Mo_ratio_relation_corrected}
\end{equation}
where $\alpha$ and $\beta$ (both greater than 0) characterize the effects of microscopic mixing and the kinetic energy respectively. Here, the effect of microscopic mixing is absorbed by the dependence on $M_{\rm O}$ because the stellar evolution models show that the occurrence rate of shell-merger during the Si burning stage is dependent on the progenitor mass, and more massive stars would have a higher chance of calcium pollution (\citealt{collins18,dessart20}).

\subsubsection{$M_{\rm CO}$-$E_{\rm K}$ relation of SNe IIb/Ib}
\begin{figure}[hbt]
\epsscale{1.05}
\plotone{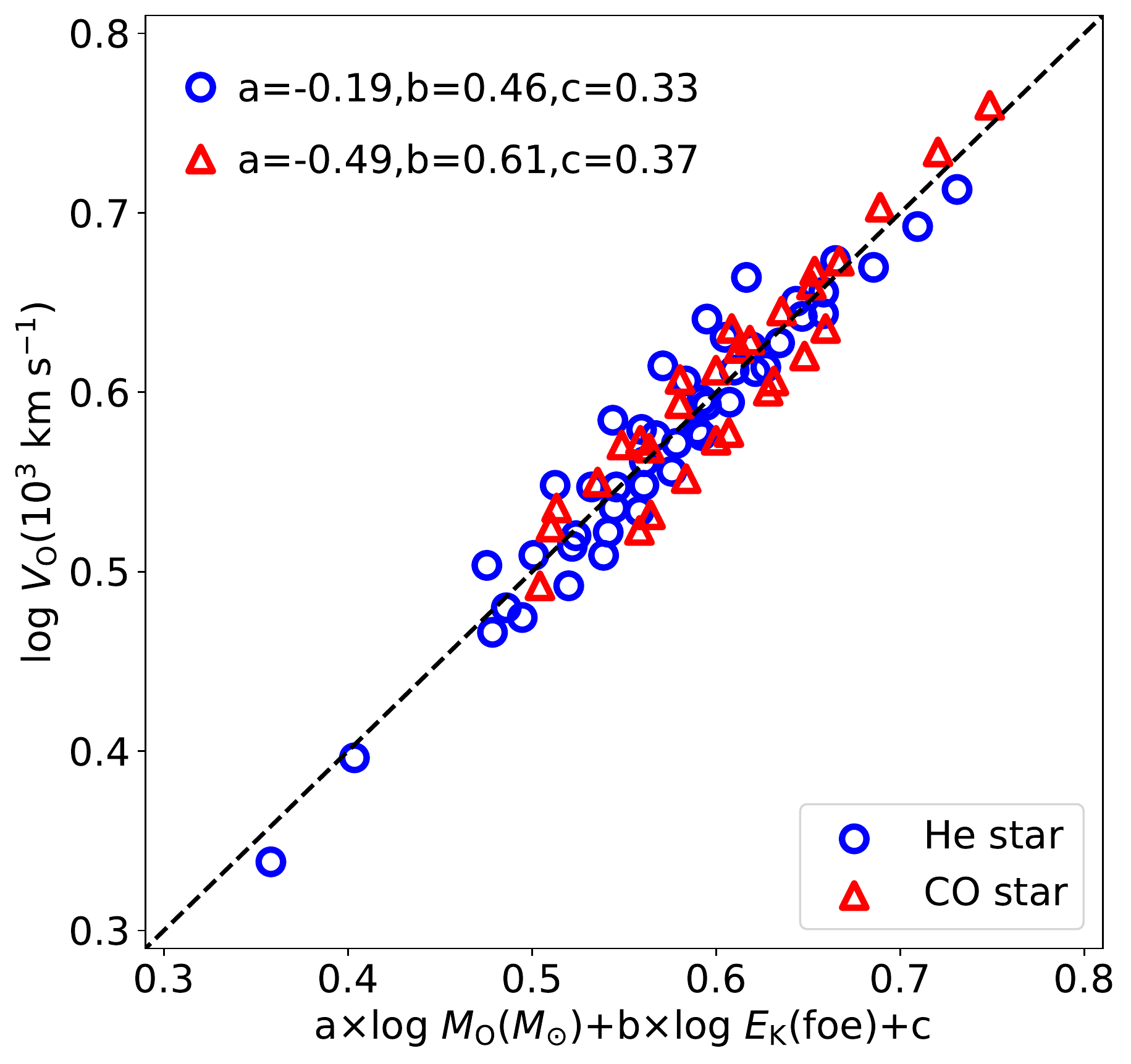}
\centering
\caption{The linear regression to the model line width $V_{\rm O}(M_{\rm O}, E_{\rm K})$ as function of oxygen mass $M_{\rm O}$ and kinetic energy $E_{\rm K}$. The helium star and the CO star models are labeled by different colors and markers. The black dashed line is one-to-one correspondence.}
\label{fig:regression}
\end{figure}

To examine the effects of $E_{\rm K}$ and microscopic mixing on the $M_{\rm CO}$-$E_{\rm K}$ relation, we first need to derive the scaling relations between the observables and the models. For the He star models with $f=0.368$, the measured line width is determined by $M_{\rm O}$ and $E_{\rm K}$, and the linear regression gives
\begin{equation}
\begin{split}
    {\rm log}\frac{V_{\rm O}}{{\rm 10^3 ~km~s^{-1}}}=(-0.20\pm 0.01)\times{\rm log}\frac{M_{\rm O}}{M_{\odot}} + \\
    (0.46\pm0.02)\times {\rm log}\frac{E_{\rm K}}{\rm foe} + (0.33\pm 0.01).
\label{eq:He_O_E_relation}
\end{split}
\end{equation}
as shown in Figure \ref{fig:regression}.

The relation between the observed line width $V_{\rm obs}$ and the [O~I]/[Ca~II] can also be expressed in the form of power-law derived from the linear regression in logarithm scale: 
\begin{equation}
    {\rm log}\frac{V_{\rm Obs}}{\rm 10^3 km~s^{-1}}=(0.18\pm0.04)\times{\rm log}\frac{\rm [O~I]}{\rm [Ca~II]}+0.41\pm0.01.
\label{eq:He_obs}
\end{equation}

By combining Equations \ref{eq:O_CO_preSN_relation}, \ref{eq:Mo_ratio_relation_corrected}, \ref{eq:He_O_E_relation} and \ref{eq:He_obs}, we have $E_{\rm K}\propto M_{\rm CO}^\delta$ (including $\alpha$ and $\beta$ as unknown parameters), where
\begin{equation}
    \delta=\frac{0.63-0.31\alpha}{0.46+0.18\beta}.
\label{eq:EK_MO_relation_corrected}
\end{equation}
If $\alpha,\beta$=0 (in this case, Equation~\ref{eq:Mo_ratio_relation_corrected} recovers Equation~\ref{eq:model_to_obs1}, where [O~I]/[Ca~II]$\propto M_{\rm O}^{0.9}$), then $\delta$=1.37, which is similar to Equation~\ref{eq:Ek_O_relation_He_only} as expected. With Equation~\ref{eq:EK_MO_relation_corrected}, we can investigate how the scaling index $\delta$ of $M_{\rm CO}$-$E_{\rm K}$ relation is affected by the effect of $E_{\rm K}$ and the microscopic mixing (characterized by the parameters $\beta$ and $\alpha$ respectively).

We first consider the effect of $E_{\rm K}$ on the [O~I]/[Ca~II] ratio. In the nebular models of \citet{fransson89}, [O~I]/[Ca~II] scales as $E_{\rm K}^{-0.5}$. In this case ($\beta$=0.5 and $\alpha$=0), we have $\delta$=1.14. Still, this would require $E_{\rm K}$ tightly correlated with $M_{\rm CO}$, although the dependence is slightly weaker than Equation \ref{eq:Ek_O_relation_He_only}.

Lacking a large grid of nebular spectra models with different degrees of microscopic mixing, it is difficult to derive the exact value of $\alpha$. However, its range can be roughly constrained from observation. If $\alpha$ is larger than 0.9, then according to Equation \ref{eq:Mo_ratio_relation_corrected}, the [O~I]/[Ca~II] ratio will be anti-correlated with the progenitor oxygen mass $M_{\rm O}$.
However, \citet{fang19} find a correlation between the [O~I]/[Ca~II] ratio and the light curve width. The light curve width measures the diffusion time scales of the photons, which is the independent measurement of the ejecta mass (as the representation of the progenitor mass). This correlation is justified by the \cite{karame2022}: their sample of SESNe with broad light curve have distinctly larger [O~I]/[Ca~II]. The [O~I]/[Ca~II] ratio is not heavily affected by the microscopic mixing (otherwise this correlation would not be expected), but the oxygen content in the ejecta should be the dominating factor, with larger [O~I]/[Ca~II] implying a more massive CO core.

Although the degree of the pre-SN calcium pollution is difficult to be inferred from the current observation, its effect on [O~I]/[Ca~II] is probably not very strong, and $\alpha$ can not be too large, or at least should be smaller than 0.9. With this constraint, $\delta$\textgreater0.76 if $\beta$=0, according to Equation~\ref{eq:EK_MO_relation_corrected}. 

In the most extreme case where $\alpha$=0.9 and $\beta$=0.5, we have $\delta$=0.64. In conclusion, even the effects of kinetic energy and calcium pollution are taken into account, we would still expect a positive correlation between $E_{\rm K}$ and $M_{\rm CO}$ to explain the observed correlation in Figure \ref{fig:main}. However, to derive the exact relation between $E_{\rm K}$ and $M_{\rm CO}$ based on the correlation between [O~I]/[Ca~II] and [O~I] width, sophisticated models that can constrain both $\alpha$ and $\beta$ are needed.

\subsubsection{$M_{\rm CO}$-$E_{\rm K}$ relation of SNe Ic/Ic-BL}
Similar to the practice of the previous section, for the CO star models, the relation between the model line width $V_{\rm O}$, kinetic energy $E_{\rm K}$ and model oxygen mass $M_{\rm O}$ is given by:
\begin{equation}
\begin{split}
    {\rm log}\frac{V_{\rm O}}{{\rm 10^3 ~km~s^{-1}}}=(-0.49\pm 0.05)\times{\rm log}\frac{M_{\rm O}}{M_{\odot}} + \\
    (0.61\pm0.04)\times {\rm log}\frac{E_{\rm K}}{\rm foe} + (0.37\pm 0.02),
\label{eq:CO_O_E_relation}
\end{split}
\end{equation}
as shown in Figure \ref{fig:regression}. Also, the relation between the observed line width $V_{\rm Obs}$ and the [O~I]/[Ca~II] ratio is given by 
\begin{equation}
    {\rm log}\frac{V_{\rm Obs}}{\rm 10^3 km~s^{-1}}=(0.04\pm0.05)\times{\rm log}\frac{\rm [O~I]}{\rm [Ca~II]}+0.48\pm0.02.
\label{eq:CO_obs}
\end{equation}

For SNe Ic + Ic-BL, the combination of Equations~\ref{eq:O_CO_preSN_relation}, \ref{eq:Mo_ratio_relation_corrected}, \ref{eq:CO_O_E_relation} and \ref{eq:CO_obs} gives 

\begin{equation}
    \delta = \frac{0.89 -0.07\alpha}{0.61 + 0.04\beta}.
\label{eq:EK_MO_relation_corrected_CO}
\end{equation}
If $\alpha,~\beta$=0, $\delta$=1.46, which is consistent with Equation \ref{eq:Ek_O_relation_CO_only} as expected. Unlike the helium-rich SNe, the effects of kinetic energy ($\beta$) and the level of microscopic mixing ($\alpha$) on $\delta$ is very weak. In the most extreme case where $\alpha$=0.9 and $\beta$=0.5, we still have $\delta$=1.31.

The derivation of Equation \ref{eq:EK_MO_relation_corrected_CO} is based on the assumption that the CO star models follow the same $M_{\rm O}$-[O~I]/[Ca~II] scaling relation as the helium star models (Equation \ref{eq:model_to_obs1} or \ref{eq:Mo_ratio_relation_corrected}). However, as noted above, these relations are not necessarily valid for the CO star models. Observationally, for SNe Ic/Ic-BL, the dependence of the [O~I] width on [O~I]/[Ca~II] is weak. By applying Equation~\ref{eq:O_CO_preSN_relation} and \ref{eq:CO_O_E_relation} with $V_{\rm O}$ fixed to be a constant (Figure \ref{fig:main}) and $\alpha$,$\beta$=0, we have
\begin{equation}
    E_{\rm K}\propto M_{\rm CO}^{1.40}.
\label{eq:Ic_MO_EK_relation_scaling}
\end{equation}
For the helium-deficient SNe, although currently we lack consistent SNe Ic nebular spectra models to constrain the relation between $M_{\rm O}$ and [O~I]/[Ca~II], still the power index $\delta$ derived from the simple scaling analysis (Equation \ref{eq:Ic_MO_EK_relation_scaling}) is consistent with that of the helium-rich models, which again suggests the SESNe share the same explosion mechanism. 
 
\subsection{Effect of macroscopic mixing}
Large-scale material mixing (macroscopic mixing) in core-collapse SN ejecta would occur due to the instability which likely arises during the explosion. It is expected to take place at the interface between the CO core and the He-rich layer, bring up the material in the CO core to the outer region. If $^{56}$Ni and oxygen are mixed into the outer region (therefore with faster expansion velocity according to the assumption of homologous expansion), the line width will increase based on Equation~\ref{eq:line_profile}. In particular, the mixing of the radioactive $^{56}$Ni strongly affects the electromagnetic properties and the thermal conditions. The line width is therefore affected by the interplay of these factors even the pre-explosion structure and the kinetic energy $E_{\rm K}$ are fixed.
In this section we will investigate whether the degree of mixing can account for the observed large scatter in [O~I] width and affect the $M_{\rm CO}$-$E_{\rm K}$ relation.

\begin{figure}[hbt]
\epsscale{1.05}
\plotone{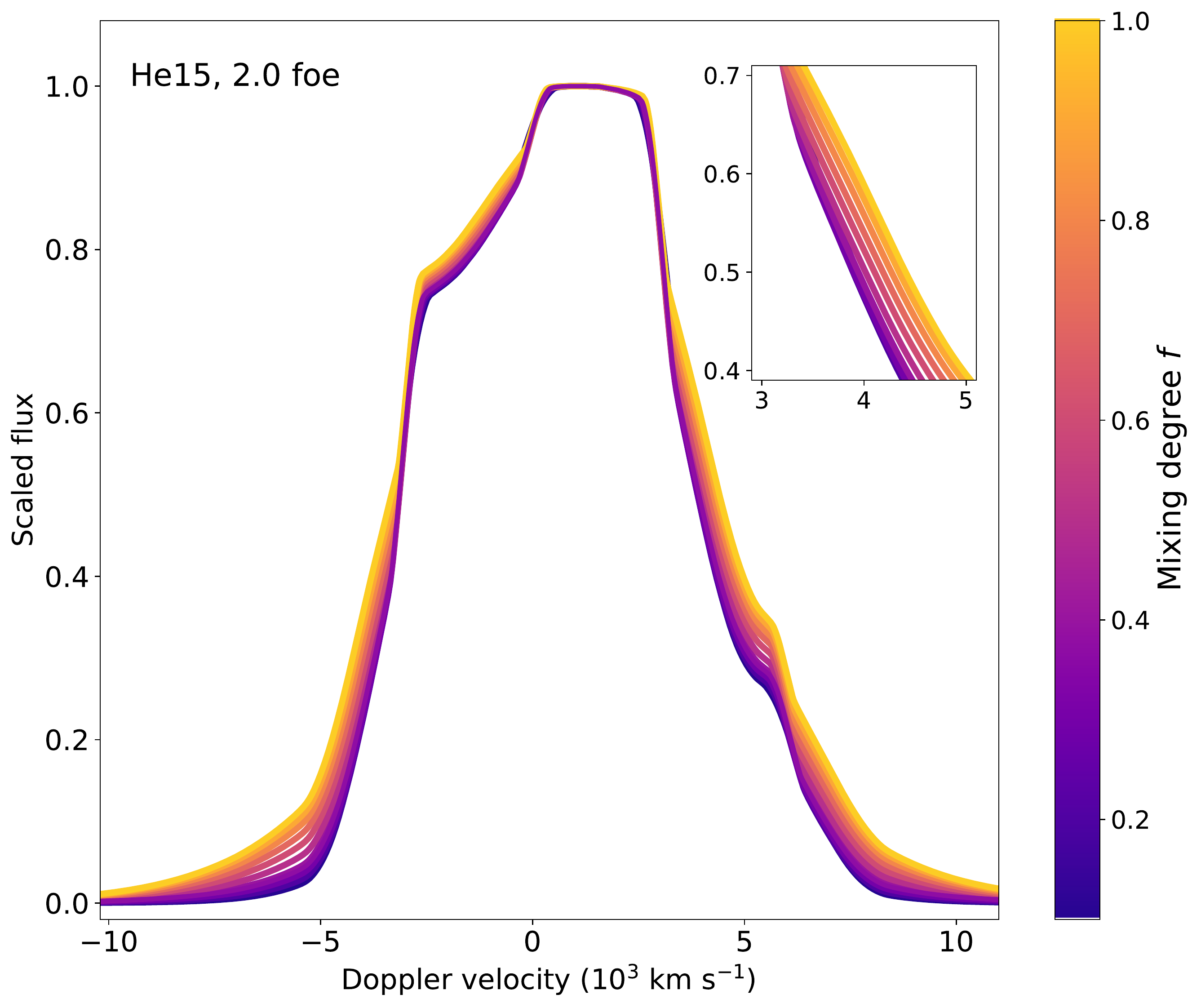}
\centering
\caption{The [O~I] profiles of He15 model ($E_{\rm K}$ = 2.0 foe) with different degrees of macroscopic mixing, labeled by the colorbar.}
\label{fig:mix_effect_line}
\end{figure}

\begin{figure}[hbt]
\epsscale{1.05}
\plotone{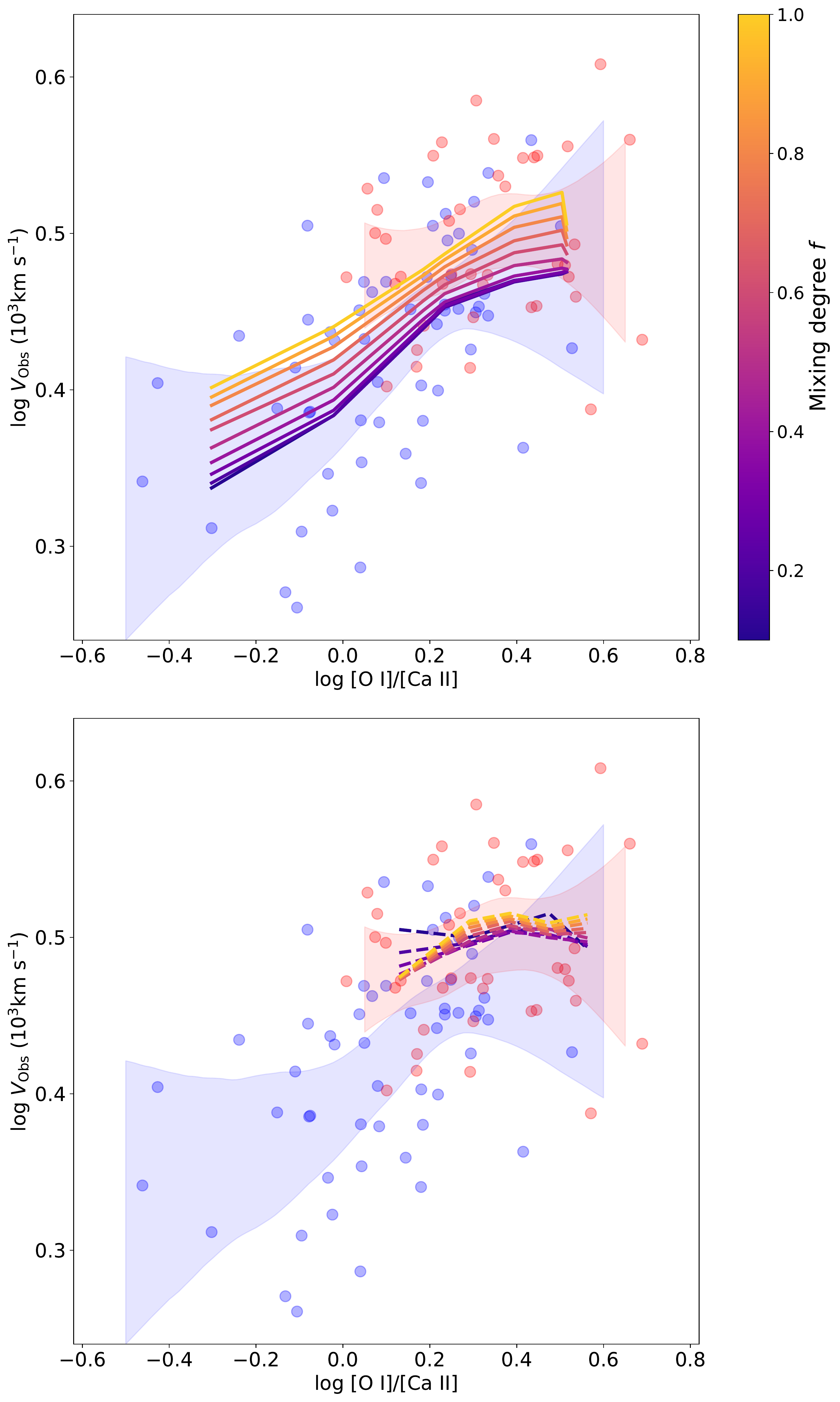}
\centering
\caption{The [O~I]/[Ca~II]-line width track of models with different degrees of macroscopic mixing (labeled by the colorbar) while the $M_{\rm CO}$-$E_{\rm K}$ relation is fixed (Table \ref{tab:kinetic_energy}). The observed [O~I]/[Ca~II]-[O~I] width relations of SNe IIb/Ib and SNe Ic/Ic-BL are illustrated as the shaded regions for comparison. $Upper~panel$: The tracks of the He star models are labeled by the solid lines; $Lower~panel$: The tracks of the CO star models are labeled by the dashed lines.}
\label{fig:mix_effect}
\end{figure}

Using the mixing scheme introduced in \S2.2, we artificially vary the degree of mixing $f$ from 0.1 to 1.0 (being fully mixed), and calculate the [O~I] profiles for different progenitor models with different kinetic energies. The [O~I] profiles of He15 model ($E_{\rm K}$=2.0 foe) calculated with different $f$ values are shown in Figure~\ref{fig:mix_effect_line} as examples. The increase of $f$ indeed leads to larger line width. To investigate the effect of $f$ on the observed line width, we calculate $V_{\rm Obs}$ for each of the progenitor model with $f$ varied and the $M_{\rm CO}$-$E_{\rm K}$ relation kept fixed (Table \ref{tab:kinetic_energy}). The expected [O~I]/[Ca~II]-[O~I] width relations are shown in Figure \ref{fig:mix_effect} for different $f$ values. For the same explosion of the same He-rich progenitor, the different degrees of large scale material mixing indeed create the scatter in line width, and can fully account for the observed scatter (the blue shaded region in Figure \ref{fig:mix_effect}). However, for the CO star models, the effect of large scale mixing on the line width is negligible. Unlike the He star models, where the material in the CO core are dredged-up to the outer region, for the CO star models, the mixing process will bring the O-rich material down to the inner region and the average velocity is reduced (lower panel in Figure \ref{fig:ni_mix}). This effect is canceled out with the dredge-up of the radioactive $^{56}$Ni.

By studying the color evolution of early phase light curves, \citet{yoon19} find evidence that the ejecta of SNe Ic is fully mixed, while for SNe IIb/Ib, the radiative $^{56}$Ni is only mildly mixed into the helium-rich envelope. This is also supported by the study of early photospheric velocity evolution; \citet{moriya20} find the helium star model can explain the photospheric velocity evolution of type Ib SN 2007Y, if the mixing process penetrates up to the middle of the ejecta. These investigations suggest the degree of mixing is possibly related to the properties of the progenitor. In this work, we have assumed the models have the same degree of mixing ($f$=0.368). If $f$ is mass-dependent, for example, in the case where more massive progenitors would lead to a larger value of $f$, the required kinetic energy will decrease to reproduce the fixed observed velocity; this reduces the slope in Equation~\ref{eq:Ek_O_relation_He_plus_CO}. We further investigate whether the change of the degree of mixing $f$ will affect the $M_{\rm CO}$-$E_{\rm K}$ relation.

\begin{figure}[hbt]
\epsscale{1.05}
\plotone{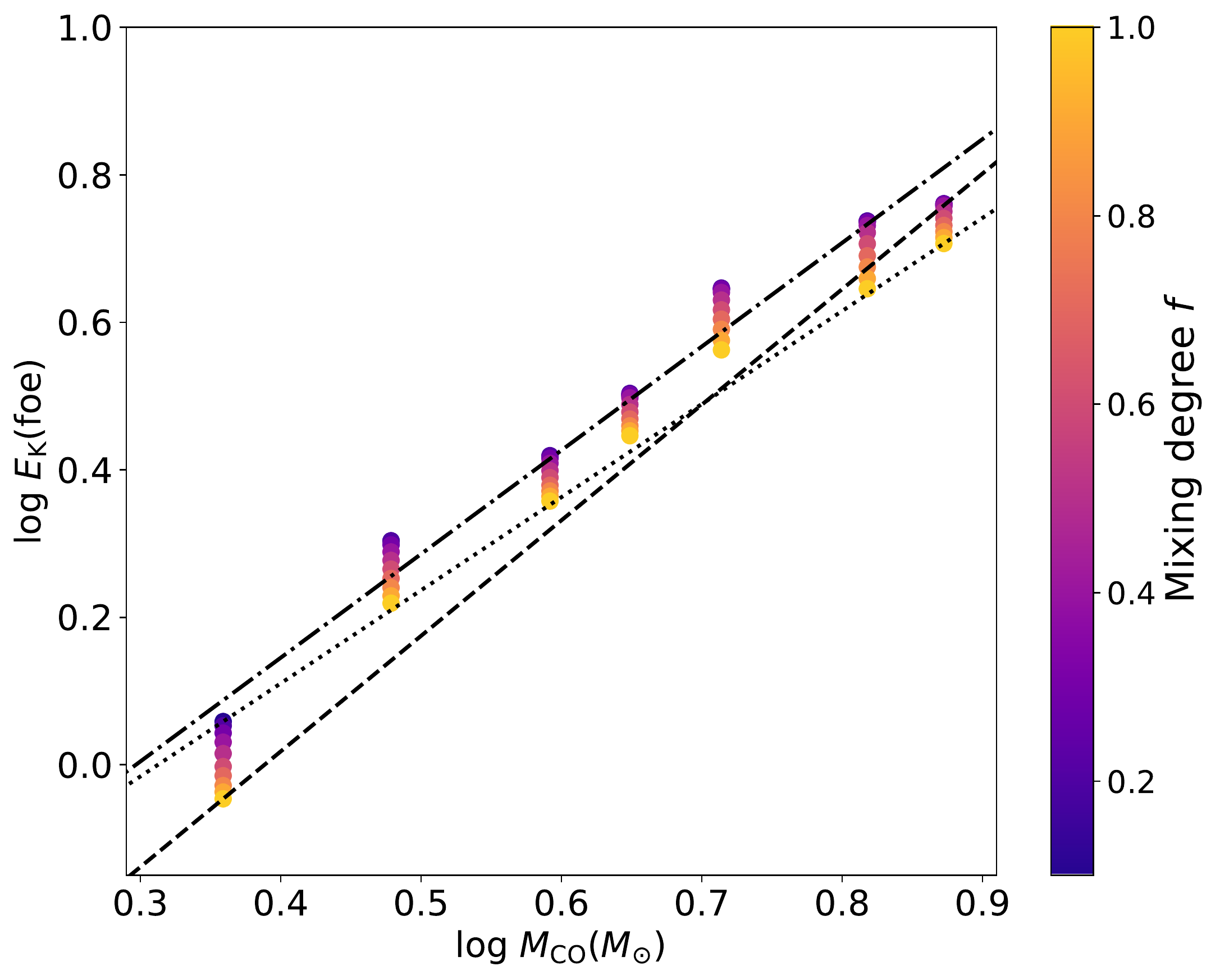}
\centering
\caption{The $M_{\rm CO}$-$E_{\rm K}$ relation required to produce the observed [O~I][Ca~II]-line width relation for the helium star models with different degrees of macroscopic mixing, labeled by the colorbar. The dashed line is the $M_{\rm CO}$-$E_{\rm K}$ relation if the mixing degree $f$ is anti-correlated with the progenitor mass; The dotted line is the $M_{\rm CO}$-$E_{\rm K}$ relation if the mixing degree $f$ is correlated with the progenitor mass. The dot-dash line is the $M_{\rm CO}$-$E_{\rm K}$ relation with $f$ fixed to 0.368 (lower panel of Figure \ref{fig:main}) for comparison.}
\label{fig:mix_effect_MO_EK}
\end{figure}

Similarly to the process in \S 3.3, we derived $E_{\rm K}$ for each progenitor model with different degrees of macroscopic mixing $f$ based on the observed line width. We consider two cases (1) $f$ is positively correlated with progenitor mass, i.e., the ejecta of a more massive star is more thoroughly mixed, with $f$ = 0.1 for He13 model and $f$ = 1.0 for He28 model; (2) $f$ is anti-correlated with progenitor mass, with $f$ = 1.0 for He13 model and $f$ = 0.1 for He28 model. The results are shown in Figure \ref{fig:mix_effect_MO_EK}, with models with different $f$ labeled by the colorbar. For case (1), we have $E_{\rm K}\propto M_{\rm CO}^{1.26}$. For case (2), the dependency increases to $E_{\rm K}\propto M_{\rm CO}^{1.58}$, as illustrated by the dotted line and dashed line in Figure \ref{fig:mix_effect_MO_EK} respectively. In conclusion, even the relation between the mixing degree and the progenitor is unknown to the current knowledge, the $M_{\rm CO}$-$E_{\rm K}$ relation will not be significantly affected.

\section{Comparison with early phase observations}
The relation between the properties of the progenitor and the kinetic energy has long been studied. However, most of the previous investigations focus on the early phase (or photospheric phase) observation (\citealt{lyman16,taddia18}). The width of the light curve and the photospheric velocity estimated from early phase spectra are used to derive the mass of the ejecta and the kinetic energy of the explosion, based on the model of \citet{arnett82}, where the ejecta is predominantly powered by the decay of the radioactive $^{56}$Ni and $^{56}$Co.

During the photospheric phase, the ejecta is optically thick. Instead of scanning through the ejecta, the information brought by analyzing the early phase observational data is limited to the properties of the outer region. The bulk properties of the ejecta are then estimated from the extrapolation inward based on several simplified assumptions (\citealt{arnett82,valenti08,cano13,lyman16,taddia18}). The observations at the photospheric phase and nebular phase are indeed tracing different regions of the ejecta, therefore it is important to compare the results derived from the two observations to clarify the validity of the assumptions.

The first step in the investigation on this topic is to connect the early phase/nebular phase observables with the models. In this section, we employ the results of \citet{lyman16} and \citet{taddia18}, which derive the ejecta mass $M_{\rm ejecta}$ and the kinetic energy $E_{\rm K}$ from the early-phase multi-band light curve of large samples of SESNe, based on the Arnett model and the radiation hydrodynamic model respectively. 

The ejecta mass estimated from the early phase observables are transformed to the pre-SN mass by
\begin{equation}
    M_{\rm pre-SN} = M_{\rm ejecta} + 1.4M_{\odot},
\label{eq:ejecta_preSN}
\end{equation}
assuming that the remnant of the explosion is fixed to 1.4$M_{\odot}$.
For the He star models, the pre-SN mass is further transformed to the CO core mass $M_{\rm CO}$ by 

\begin{equation}
    {\rm log}\frac{M_{\rm CO}}{M_{\odot}} = (1.245\pm 0.008)\times{\rm log}\frac{M_{\rm pre-SN}}{M_{\odot}}-0.366\pm 0.006.\\
\label{eq:O_preSN_He}
\end{equation}
Similarly, for the CO core models, we have
\begin{equation}
    {\rm log}\frac{M_{\rm CO}}{M_{\odot}} = (1.013\pm 0.008)\times{\rm log}\frac{M_{\rm pre-SN}}{M_{\odot}}-0.018\pm 0.006.\\
\label{eq:O_preSN_CO}
\end{equation}

We First anchor the absolute scale of the ejecta mass from the early phase analysis of \citet{lyman16}. The ejecta mass of iPTF 13bvn derived from the Arnett model is multiplied by a constant to match with the He13 model, which gives
\begin{equation}
    {\rm log}\frac{M_{\rm ejecta, model}}{M_{\odot}} ={\rm log}\frac{M_{\rm ejecta, LC}}{M_{\odot}} + 0.15.
\label{eq:ejecta_neb_LC}  
\end{equation}
Here, $M_{\rm ejecta, model}$ and $M_{\rm ejecta, LC}$ are the ejecta mass of the progenitor model and the ejecta mass estimated from the early phase light curve respectively. For the sample of \citet{taddia18}, we directly apply their $M_{\rm ejecta}$, as it was estimated based on the radiation hydrodynamic simulation. The ejecta mass is further transformed to the CO core mass using Equations~\ref{eq:ejecta_preSN}, \ref{eq:O_preSN_He}, \ref{eq:O_preSN_CO} and \ref{eq:ejecta_neb_LC}. The $M_{\rm CO}$ are then compared with the kinetic energies derived from the early phase light curve. 

The $M_{\rm CO}$-$E_{\rm K}$ relations based on the early-phase analyses from \citet{lyman16} and \citet{taddia18} are plotted in the upper and lower panels of Figure \ref{fig:early_correlation} respectively. The helium-rich SNe (IIb + Ib) and the helium-deficient SNe (Ic + Ic-BL) are labeled by different colors and markers. The $M_{\rm CO}$-$E_{\rm K}$ relation derived from the nebular spectra (lower panel of Figure \ref{fig:main}) is also plotted for comparison.

\subsection{Comparison with \citet{lyman16}}
$M_{\rm ejecta}$ and $E_{\rm K}$ of the \citet{lyman16} sample are derived based on the Arnett model with several simplified assumptions, for which the readers may refer to \citet{arnett82} and \citet{lyman16} for more details.

For the \citet{lyman16} sample, the linear regressions to SNe IIb+Ib and SNe Ic+Ic-BL give
\begin{equation}
    E_{\rm K} \propto M_{\rm CO}^{1.31\pm 0.18}\\
\end{equation}
and
\begin{equation}
    E_{\rm K} \propto M_{\rm CO}^{1.18\pm 0.33}\\
\end{equation}
respectively. 
If the linear regression is performed to the full sample, we have
\begin{equation}
    E_{\rm K} \propto M_{\rm CO}^{1.36\pm 0.16}\\
\end{equation}

The slope of the $M_{\rm CO}$-$E_{\rm K}$ relation of SNe IIb+Ib is consistent with the one derived from the nebular phase observation. The consistency between the anlayses using the early phase and nebular phase observables further suggests the effects of $E_{\rm K}$ and the degree of microscopic mixing on [O~I]/[Ca~II] is probably not very strong. To be more specific, we now look into Equation~\ref{eq:EK_MO_relation_corrected}. To match with the result from the nebular phase observation, with $\delta$=1.31 derived from the early phase observation, the values of $\alpha$ and $\beta$ are constrained by
\begin{equation}
    0.31\alpha+0.24\beta=0.03,
\label{eq:constrain_alpha_beta_He_L16}
\end{equation}
therefore $\alpha\textless$0.10 and $\beta\textless$0.13 ($\alpha, \beta \textgreater$0; see discussions in \S 4.1).

\begin{figure}[hbt]
\epsscale{1.05}
\plotone{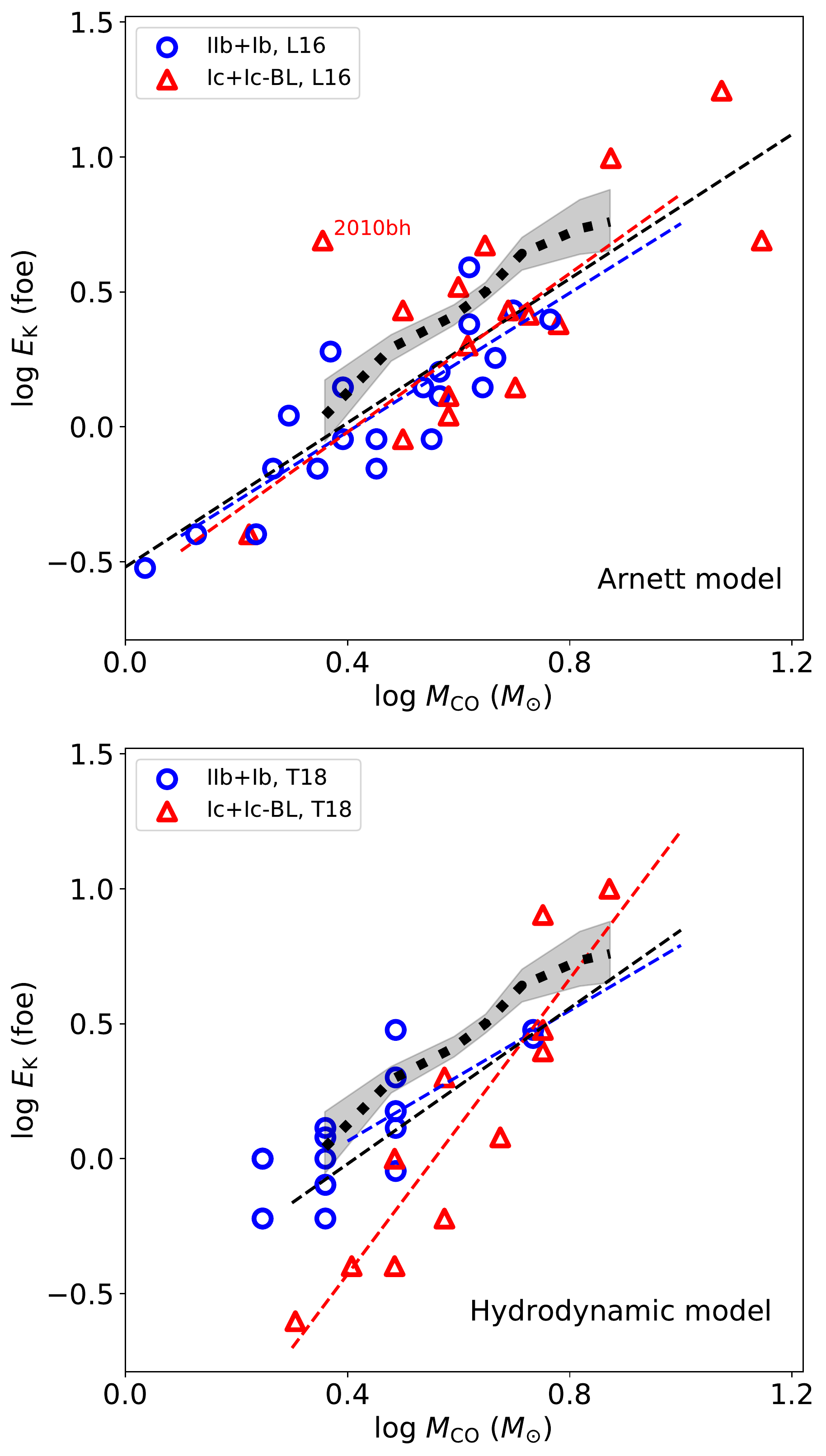}
\centering
\caption{The $M_{\rm CO}$-$E_{\rm K}$ relation derived using the early phase observables. The scatter points are individual objects, with SNe IIb + Ib labeled by blue circles and SNe Ic + Ic-BL labeled by red triangles. The blue and red dashed lines are the linear regressions to the helium-rich and deficient SNe respectively. The black dashed lines are the linear regressions to the full sample. The black dotted line is the result derived from the nebular spectrum analysis, with the shaded area showing the 95\% confidence level (lower panel of Figure \ref{fig:main}). $Upper~panel$: The relation based on the early phase observation from \citet{lyman16}, with $M_{\rm ejecta}$ and $E_{\rm K}$ estimated from the Arnett model; $Lower~panel$: The relation based on \citet{taddia18}, with $M_{\rm ejecta}$ and $E_{\rm K}$ estimated from the hydrodynamic model.}
\label{fig:early_correlation}
\end{figure}

For the He-deficient SNe, the power-law index $\delta$ derived from the early phase observation is smaller than the one derived from nebular phase (Equation \ref{eq:Ic_MO_EK_relation_scaling}), but still the two relations are consistent within uncertainty. Further, if the possible outlier SN 2010bh is excluded (as labeled out in the upper panel of Figure \ref{fig:early_correlation}), the linear regression gives
\begin{equation}
    E_{\rm K} \propto M_{\rm CO}^{1.47\pm 0.29}.\\
\end{equation}

In conclusion, for SNe IIb/Ib and SNe Ic/Ic-BL, the $M_{\rm CO}$-$E_{\rm K}$ relations from both the early phase and nebular phase observations are consistent.

\subsection{Comparison with \citet{taddia18}}
Instead of using Arnett model, $M_{\rm ejecta}$ and $E_{\rm K}$ of the \citet{taddia18} sample is derived based on the radiation hydrodynamic model. The light curve of the SNe in the sample is compared with the simulated light curves, which are generated by exploding a series of helium star models with different progenitor masses by a range of the kinetic energy. The ejecta mass $M_{\rm ejecta}$ of the \citet{taddia18} sample is transformed to the CO core mass $M_{\rm CO}$ via Equations \ref{eq:ejecta_preSN}, \ref{eq:O_preSN_He}, and \ref{eq:O_preSN_CO}.

The linear regressions to SNe IIb+Ib and SNe Ic+Ic-BL of the \citet{taddia18} sample give
\begin{equation}
    E_{\rm K} \propto M_{\rm CO}^{1.23\pm 0.22}\\
\label{eq:EK_MO_He_T18}
\end{equation}
and
\begin{equation}
    E_{\rm K} \propto M_{\rm CO}^{2.74\pm 0.39}\\
\label{eq:EK_MO_CO_T18}
\end{equation}
respectively. The $M_{\rm CO}$-$E_{\rm K}$ relation of SNe IIb+Ib derived based on early phase observation is consistent with the one from the nebular phase observation within uncertainty. Similar to the analysis to the \citet{lyman16} sample, Equation \ref{eq:EK_MO_He_T18} constrains the value of $\alpha$ and $\beta$ through
\begin{equation}
    0.31\alpha+0.22\beta=0.06,
\label{eq:constrain_alpha_beta_He_T18}
\end{equation}
and $\alpha\textless$0.21 and $\beta\textless$0.29, i.e., the effects of $E_{\rm K}$ and microscopic mixing on [O~I]/[Ca~II] are negligible, which is similar with the case of \citet{lyman16} sample.

However, for the SNe Ic/Ic-BL sample, the slope of Equation~\ref{eq:EK_MO_CO_T18} is much larger than the ones derived from the nebular analysis (Equation~\ref{eq:Ek_O_relation_CO_only}) and the sample of \citet{lyman16}. This is possibly because \citet{taddia18} estimate $E_{\rm K}$ and $M_{\rm ejecta}$ of the helium-deficient SNe by comparing their observed light curves with the simulated light curves of the helium-rich star models. This potentially introduces a systematic offset in $E_{\rm K}$ and $M_{\rm ejecta}$, which in turn affects the slope of Equation \ref{eq:EK_MO_CO_T18}.

\section{Summary}
Based on the large nebular spectra sample of stripped-envelope core-collapse supernovae, \citet{fang22} found a correlation between [O~I]/[Ca~II] (which measures the progenitor mass) and [O~I] width (which measures the expansion velocity of the O-rich material). This work aims to explain this correlation from a theoretical aspect.

One-dimensional simulations of massive-star evolution from 13 to 28$M_{\odot}$, with the hydrogen envelope or the helium-rich layer stripped, are carried out by \texttt{MESA}. When the massive stars evolve to the time of core-collapse, they are used as the input models for \texttt{SNEC}, and further exploded as CCSNe by injecting different amount of the kinetic energy (1$\sim$10$\times 10^{51}$ erg) into the central regions. The oxygen mass of the model is transformed to the [O~I]/[Ca~II] ratio by assuming the scaling relation derived from the nebular SN IIb models of \citet{jerkstrand15}. The velocity of the O-rich materials as weighted by the local $\gamma$-ray deposition rate is connected to the observed [O~I] width. The analysis in this work suggests the following to produce the correlation between the [O~I]/[Ca~II] ratio and the [O~I] width; the kinetic energy of the explosion should correlate with the CO core mass of the ejecta, and scales as $E_{\rm K}\propto M_{\rm CO}^{1.41}$. Further, SNe Ic/Ic-BL follow almost the same $M_{\rm CO}$-$E_{\rm K}$ relation as SNe IIb/Ib, i.e., $E_{\rm K}\propto M_{\rm CO}^{1.34}$, which suggests the helium-rich and helium-deficient SNe likely share the same explosion mechanism.

However, the above inferences are made based on several simplified assumptions and empirical relations (for example, we adopt a specific model sequence for SNe IIb by \citealt{jerkstrand15} for the conversion between the [O~I]/[Ca~II] ratio and $M_{\rm O}$). Lacking consistent nebular model spectra of SESNe exploded by a large range of the kinetic energy, it is difficult to estimate the accuracy of the $M_{\rm CO}$-$E_{\rm K}$ relation derived from the method presented in this work.
We have discussed several factors that would possibly affect the result. With the scaling analysis, we conclude that the power-law index of the $M_{\rm CO}$-$E_{\rm K}$ relation of the helium-rich SNe is affected by the dependence of the [O~I]/[Ca~II] ratio on $E_{\rm K}$ and the degree of microscopic mixing. However, the power-law index of the $M_{\rm CO}$-$E_{\rm K}$ relation is insensitive to such dependence, especially for the helium-deficient SNe. Further, the macroscopic mixing potentially developed during the explosion will bring the material in the CO core up to outer region, increasing the average velocity of the O-rich material and the [O~I] width. Different degrees of macroscopic mixing can create the scatter in the observed line widths.

The derivation of an accurate $M_{\rm CO}$-$E_{\rm K}$ relation is associated with these complicated physical issues, therefore would require a large grid of detailed radiative-transfer modeling with the above factors taken into account. Sophisticated stellar evolution modeling is also needed to estimate the occurrence rate of the microscopic mixing of the calcium into the O-rich shell, which is caused by the shell merger developed during the advanced nuclear burning stage.

With the above caveats in mind, we compare the $M_{\rm CO}$-$E_{\rm K}$ relation derived from this work with the early phase observation of \citet{lyman16} and \citet{taddia18}. During the early phase, the ejecta is optically thick, and the observation traces the nature of the outer region. When the ejecta enters nebular phase, it becomes transparent, and the observation probes the nature of the densest region, i.e., the innermost part. The observations at different phases are thus independently constraining the natures of different regions within the ejecta. Interestingly, for the helium-rich SNe, the $M_{\rm CO}$-$E_{\rm K}$ relation derived from these two methods are in good agreement. It is largely the case for the helium-deficient SNe as well, but the situation is less clear; while the scaling we have derived for the core region agrees with the one derived from the outer region by \citet{lyman16}, the power-law index of the $M_{\rm CO}$-$E_{\rm K}$ derived from the sample of \citet{taddia18} is too steep compared with the observation of nebular phase. This is possibly because the analysis of the SNe Ic/Ic-BL in the sample of \citet{taddia18} is based on helium-rich models. We emphasize that the $M_{\rm CO}$-$E_{\rm K}$ relations derived for the outer region (by the early-phase analysis) and for the innermost region (by the late-phase analysis) do not have to agree, as different regions are probed.

In this work, we present a method to investigate the relation between the progenitor mass and the kinetic energy of the explosion through the nebular-phase observation. Although this method suffers from the lack of consistent nebular spectra models, it can serve as a cross-reference of the ejecta properties inferred from the early-phase observation, which is frequently adopted in the previous literature. The combined analysis of the observational data in the early and late phases provides us the chance to scan through the full ejecta from the outermost region to the dense inner part. Not only the consistency, but also the inconsistency of the two methods, is important to investigate the completeness of the current assumptions on the explosion process, which is crucial to reveal the explosion mechanism of core-collapse supernovae.

\begin{acknowledgements}
The authors would like to thank the anonymous reviewer for the comments that helped to improve the manuscript. Q.F. acknowledges support by Japan Society for the Promotion of Science (JSPS) KAKENHI Grant (20J23342). K.M. acknowledges support by JSPS KAKENHI Grant (18H05223, 20H00174, 20H04737). 
\end{acknowledgements}

\software{$\texttt{MESA}$ \citep{paxton11, paxton13, paxton15, paxton18, paxton19}; $\texttt{SNEC}$ \citep{snec15}; SciPy \citep{scipy}; NumPy \citep{numpy}; Astropy \citep{astropy13,astropy18}; Matplotlib \citep{matplotlib}}

{}
\end{CJK*}

\clearpage
\newpage

\begin{thebibliography}{}
\bibitem[Arnett(1982)]{arnett82} Arnett, W.~D.\ 1982, \apj, 253, 785. doi:10.1086/159681

\bibitem[Astropy Collaboration et al.(2013)]{astropy13} Astropy Collaboration, Robitaille, T.~P., Tollerud, E.~J., et al.\ 2013, \aap, 558, A33. doi:10.1051/0004-6361/201322068

\bibitem[Astropy Collaboration et al.(2018)]{astropy18} Astropy Collaboration, Price-Whelan, A.~M., Sip{\H{o}}cz, B.~M., et al.\ 2018, \aj, 156, 123. doi:10.3847/1538-3881/aabc4f

\bibitem[Bersten et al.(2013)]{bersten13} Bersten, M.~C., Tanaka, M., Tominaga, N., et al.\ 2013, \apj, 767, 143. doi:10.1088/0004-637X/767/2/143


\bibitem[Bersten et al.(2014)]{bersten14} Bersten, M.~C., Benvenuto, O.~G., Folatelli, G., et al.\ 2014, \aj, 148, 68. doi:10.1088/0004-6256/148/4/68

\bibitem[Cano(2013)]{cano13} Cano, Z.\ 2013, \mnras, 434, 1098. doi:10.1093/mnras/stt1048

\bibitem[Cano et al.(2014)]{cano14} Cano, Z., Maeda, K., \& Schulze, S.\ 2014, \mnras, 438, 2924. doi:10.1093/mnras/stt2400

\bibitem[Cao et al.(2013)]{cao13} Cao, Y., Kasliwal, M.~M., Arcavi, I., et al.\ 2013, \apjl, 775, L7. doi:10.1088/2041-8205/775/1/L7

\bibitem[Collins et al.(2018)]{collins18} Collins, C., M{\"u}ller, B., \& Heger, A.\ 2018, \mnras, 473, 1695. doi:10.1093/mnras/stx2470

\bibitem[Dessart et al.(2011)]{dessart11} Dessart, L., Hillier, D.~J., Livne, E., et al.\ 2011, \mnras, 414, 2985. doi:10.1111/j.1365-2966.2011.18598.x

\bibitem[Dessart et al.(2012)]{dessart12} Dessart, L., Hillier, D.~J., Li, C., et al.\ 2012, \mnras, 424, 2139. doi:10.1111/j.1365-2966.2012.21374.x


\bibitem[Dessart et al.(2013)]{dessart13} Dessart, L., Hillier, D.~J., Waldman, R., et al.\ 2013, \mnras, 433, 1745. doi:10.1093/mnras/stt861

\bibitem[Dessart et al.(2015)]{dessart15} Dessart, L., Hillier, D.~J., Woosley, S., et al.\ 2015, \mnras, 453, 2189. doi:10.1093/mnras/stv1747

\bibitem[Dessart et al.(2016)]{dessart16} Dessart, L., Hillier, D.~J., Woosley, S., et al.\ 2016, \mnras, 458, 1618. doi:10.1093/mnras/stw418

\bibitem[Dessart \& Hillier(2020)]{dessart20} Dessart, L. \& Hillier, D.~J.\ 2020, \aap, 642, A33. doi:10.1051/0004-6361/202038148

\bibitem[Dessart et al.(2021)]{dessart21} Dessart, L., Hillier, D.~J., Sukhbold, T., et al.\ 2021, \aap, 656, A61. doi:10.1051/0004-6361/202141927

\bibitem[Ennis et al.(2006)]{ennis06} Ennis, J.~A., Rudnick, L., Reach, W.~T., et al.\ 2006, \apj, 652, 376. doi:10.1086/508142

\bibitem[Ensman \& Woosley(1988)]{ensman88} Ensman, L.~M. \& Woosley, S.~E.\ 1988, \apj, 333, 754. doi:10.1086/166785


\bibitem[Fang \& Maeda(2018)]{fang18} Fang, Q. \& Maeda, K.\ 2018, \apj, 864, 47. doi:10.3847/1538-4357/aad096

\bibitem[Fang et al.(2019)]{fang19} Fang, Q., Maeda, K., Kuncarayakti, H., et al.\ 2019, Nature Astronomy, 3, 434. doi:10.1038/s41550-019-0710-6

\bibitem[Fang et al.(2022)]{fang22} Fang, Q., Maeda, K., Kuncarayakti, H., et al.\ 2022, \apj, 928, 151. doi:10.3847/1538-4357/ac4f60

\bibitem[Filippenko(1997)]{filippenko97} Filippenko, A.~V.\ 1997, \araa, 35, 309. doi:10.1146/annurev.astro.35.1.309

\bibitem[Fransson \& Chevalier(1989)]{fransson89} Fransson, C. \& Chevalier, R.~A.\ 1989, \apj, 343, 323. doi:10.1086/167707

\bibitem[Fremling et al.(2016)]{fremling16} Fremling, C., Sollerman, J., Taddia, F., et al.\ 2016, \aap, 593, A68. doi:10.1051/0004-6361/201628275

\bibitem[Galama et al.(1998)]{galama98} Galama, T.~J., Vreeswijk, P.~M., van Paradijs, J., et al.\ 1998, \nat, 395, 670. doi:10.1038/27150

\bibitem[Gal-Yam(2017)]{galyam17} Gal-Yam, A.\ 2017, Handbook of Supernovae, 195. doi:10.1007/978-3-319-21846-5\_35

\bibitem[Groh et al.(2013)]{groh13} Groh, J.~H., Georgy, C., \& Ekstr{\"o}m, S.\ 2013, \aap, 558, L1. doi:10.1051/0004-6361/201322369

\bibitem[Hamuy et al.(2006)]{hamuy06} Hamuy, M., Folatelli, G., Morrell, N.~I., et al.\ 2006, \pasp, 118, 2. doi:10.1086/500228

\bibitem[Harris et al.(2020)]{numpy} Harris, C.~R., Millman, K.~J., van der Walt, S.~J., et al.\ 2020, \nat, 585, 357. doi:10.1038/s41586-020-2649-2

\bibitem[Heger et al.(2003)]{heger03} Heger, A., Fryer, C.~L., Woosley, S.~E., et al.\ 2003, \apj, 591, 288. doi:10.1086/375341

\bibitem[Hjorth et al.(2003)]{hjorth03} Hjorth, J., Sollerman, J., M{\o}ller, P., et al.\ 2003, \nat, 423, 847. doi:10.1038/nature01750

\bibitem[Hunter(2007)]{matplotlib} Hunter, J.~D.\ 2007, Computing in Science and Engineering, 9, 90. doi:10.1109/MCSE.2007.55


\bibitem[Jerkstrand et al.(2015)]{jerkstrand15} Jerkstrand, A., Ergon, M., Smartt, S.~J., et al.\ 2015, \aap, 573, A12. doi:10.1051/0004-6361/201423983

\bibitem[Jerkstrand(2017)]{jerkstrand17} Jerkstrand, A.\ 2017, Handbook of Supernovae, 795. doi:10.1007/978-3-319-21846-5\_29

\bibitem[Karamehmetoglu et al.(2022)]{karame2022} Karamehmetoglu, E., Sollerman, J., Taddia, F., et al.\ 2022, arXiv:2210.09402

\bibitem[Kasen \& Woosley(2009)]{kasen09} Kasen, D. \& Woosley, S.~E.\ 2009, \apj, 703, 2205. doi:10.1088/0004-637X/703/2/2205


\bibitem[Kifonidis et al.(2003)]{mixing03} Kifonidis, K., Plewa, T., Janka, H.-T., et al.\ 2003, \aap, 408, 621. doi:10.1051/0004-6361:20030863

\bibitem[Kifonidis et al.(2006)]{mixing06} Kifonidis, K., Plewa, T., Scheck, L., et al.\ 2006, \aap, 453, 661. doi:10.1051/0004-6361:20054512

\bibitem[Kilpatrick et al.(2021)]{kilpatrick21} Kilpatrick, C.~D., Drout, M.~R., Auchettl, K., et al.\ 2021, \mnras, 504, 2073. doi:10.1093/mnras/stab838

\bibitem[Kuncarayakti et al.(2015)]{kuncarayakti15} Kuncarayakti, H., Maeda, K., Bersten, M.~C., et al.\ 2015, \aap, 579, A95. doi:10.1051/0004-6361/201425604

\bibitem[Limongi \& Chieffi(2003)]{limongi03} Limongi, M. \& Chieffi, A.\ 2003, \apj, 592, 404. doi:10.1086/375703

\bibitem[Lyman et al.(2016)]{lyman16} Lyman, J.~D., Bersier, D., James, P.~A., et al.\ 2016, \mnras, 457, 328. doi:10.1093/mnras/stv2983

\bibitem[Maeda et al.(2007)]{maeda07} Maeda, K., Kawabata, K., Tanaka, M., et al.\ 2007, \apjl, 658, L5. doi:10.1086/513564


\bibitem[Maund et al.(2011)]{maund11} Maund, J.~R., Fraser, M., Ergon, M., et al.\ 2011, \apjl, 739, L37. doi:10.1088/2041-8205/739/2/L37

\bibitem[Maurer et al.(2010)]{maurer10} Maurer, J.~I., Mazzali, P.~A., Deng, J., et al.\ 2010, \mnras, 402, 161. doi:10.1111/j.1365-2966.2009.15905.x

\bibitem[Mazzali et al.(2002)]{mazzali02} Mazzali, P.~A., Deng, J., Maeda, K., et al.\ 2002, \apjl, 572, L61. doi:10.1086/341504


\bibitem[Modjaz et al.(2008)]{modjaz08} Modjaz, M., Kewley, L., Kirshner, R.~P., et al.\ 2008, \aj, 135, 1136. doi:10.1088/0004-6256/135/4/1136

\bibitem[Modjaz et al.(2019)]{modjaz19} Modjaz, M., Guti{\'e}rrez, C.~P., \& Arcavi, I.\ 2019, Nature Astronomy, 3, 717. doi:10.1038/s41550-019-0856-2

\bibitem[Moriya et al.(2020)]{moriya20} Moriya, T.~J., Suzuki, A., Takiwaki, T., et al.\ 2020, \mnras, 497, 1619. doi:10.1093/mnras/staa2060

\bibitem[Morozova et al.(2015)]{snec15} Morozova, V., Piro, A.~L., Renzo, M., et al.\ 2015, \apj, 814, 63. doi:10.1088/0004-637X/814/1/63

\bibitem[Nakamura et al.(2001)]{nakamura01} Nakamura, T., Mazzali, P.~A., Nomoto, K., et al.\ 2001, \apj, 550, 991. doi:10.1086/319784

\bibitem[Paxton et al.(2011)]{paxton11} Paxton, B., Bildsten, L., Dotter, A., et al.\ 2011, \apjs, 192, 3. doi:10.1088/0067-0049/192/1/3

\bibitem[Paxton et al.(2013)]{paxton13} Paxton, B., Cantiello, M., Arras, P., et al.\ 2013, \apjs, 208, 4. doi:10.1088/0067-0049/208/1/4

\bibitem[Paxton et al.(2015)]{paxton15} Paxton, B., Marchant, P., Schwab, J., et al.\ 2015, \apjs, 220, 15. doi:10.1088/0067-0049/220/1/15

\bibitem[Paxton et al.(2018)]{paxton18} Paxton, B., Schwab, J., Bauer, E.~B., et al.\ 2018, \apjs, 234, 34. doi:10.3847/1538-4365/aaa5a8

\bibitem[Paxton et al.(2019)]{paxton19} Paxton, B., Smolec, R., Schwab, J., et al.\ 2019, \apjs, 243, 10. doi:10.3847/1538-4365/ab2241

\bibitem[Piro \& Nakar(2013)]{piro13} Piro, A.~L. \& Nakar, E.\ 2013, \apj, 769, 67. doi:10.1088/0004-637X/769/1/67

\bibitem[Sana et al.(2012)]{sana12} Sana, H., de Mink, S.~E., de Koter, A., et al.\ 2012, Science, 337, 444. doi:10.1126/science.1223344

\bibitem[Sauer et al.(2006)]{sauer06} Sauer, D.~N., Mazzali, P.~A., Deng, J., et al.\ 2006, \mnras, 369, 1939. doi:10.1111/j.1365-2966.2006.10438.x

\bibitem[Shigeyama \& Nomoto(1990)]{shigeyama90a} Shigeyama, T. \& Nomoto, K.\ 1990, \apj, 360, 242. doi:10.1086/169114

\bibitem[Shigeyama et al.(1990)]{shigeyama90b} Shigeyama, T., Nomoto, K., Tsujimoto, T., et al.\ 1990, \apjl, 361, L23. doi:10.1086/185818


\bibitem[Smith(2014)]{smith14} Smith, N.\ 2014, \araa, 52, 487. doi:10.1146/annurev-astro-081913-040025

\bibitem[Sun et al.(2023)]{sun23} Sun, N.-C., Maund, J.~R., \& Crowther, P.~A.\ 2023, \mnras. doi:10.1093/mnras/stad690

\bibitem[Taddia et al.(2018)]{taddia18} Taddia, F., Stritzinger, M.~D., Bersten, M., et al.\ 2018, \aap, 609, A136. doi:10.1051/0004-6361/201730844

\bibitem[Taubenberger et al.(2009)]{taubenberger09} Taubenberger, S., Valenti, S., Benetti, S., et al.\ 2009, \mnras, 397, 677. doi:10.1111/j.1365-2966.2009.15003.x

\bibitem[Teffs et al.(2020)]{teffs20} Teffs, J., Ertl, T., Mazzali, P., et al.\ 2020, \mnras, 492, 4369. doi:10.1093/mnras/staa123

\bibitem[Valenti et al.(2008)]{valenti08} Valenti, S., Benetti, S., Cappellaro, E., et al.\ 2008, \mnras, 383, 1485. doi:10.1111/j.1365-2966.2007.12647.x

\bibitem[Van Dyk et al.(2014)]{dyk14} Van Dyk, S.~D., Zheng, W., Fox, O.~D., et al.\ 2014, \aj, 147, 37. doi:10.1088/0004-6256/147/2/37

\bibitem[Virtanen et al.(2020)]{scipy} Virtanen, P., Gommers, R., Oliphant, T.~E., et al.\ 2020, Nature Methods, 17, 261. doi:10.1038/s41592-019-0686-2

\bibitem[Wongwathanarat et al.(2015)]{mixing15} Wongwathanarat, A., M{\"u}ller, E., \& Janka, H.-T.\ 2015, \aap, 577, A48. doi:10.1051/0004-6361/201425025

\bibitem[Woosley \& Weaver(1995)]{woosley95} Woosley, S.~E. \& Weaver, T.~A.\ 1995, \apjs, 101, 181. doi:10.1086/192237

\bibitem[Woosley et al.(2002)]{woosley02} Woosley, S.~E., Heger, A., \& Weaver, T.~A.\ 2002, Reviews of Modern Physics, 74, 1015. doi:10.1103/RevModPhys.74.1015

\bibitem[Woosley \& Bloom(2006)]{woosley06} Woosley, S.~E. \& Bloom, J.~S.\ 2006, \araa, 44, 507. doi:10.1146/annurev.astro.43.072103.150558

\bibitem[Yoon(2015)]{yoon15} Yoon, S.-C.\ 2015, PASA, 32, e015. doi:10.1017/pasa.2015.16

\bibitem[Yoon et al.(2019)]{yoon19} Yoon, S.-C., Chun, W., Tolstov, A., et al.\ 2019, \apj, 872, 174. doi:10.3847/1538-4357/ab0020

\end{thebibliography}
\end{document}